\documentclass[%
 aip,
 amsmath,amssymb,
reprint,%
]{revtex4-1}

\usepackage{graphicx}% Include figure files
\usepackage{dcolumn}% Align table columns on decimal point
\usepackage{bm}% bold math
\usepackage[utf8]{inputenc}
\usepackage[T1]{fontenc}
\usepackage{mathptmx}
\usepackage{etoolbox}
\usepackage{subcaption}
\usepackage{amsmath}
\usepackage{graphics}
\usepackage{dblfloatfix} 
\usepackage[arrowdel]{physics}
\usepackage{siunitx}
\graphicspath{{./Images/}}
\usepackage{comment}
\usepackage{epstopdf}
\usepackage{color, soul}
\usepackage[usenames,dvipsnames]{xcolor}

\makeatletter
\def\@email#1#2{%
 \endgroup
 \patchcmd{\titleblock@produce}
  {\frontmatter@RRAPformat}
  {\frontmatter@RRAPformat{\produce@RRAP{*#1\href{mailto:#2}{#2}}}\frontmatter@RRAPformat}
  {}{}
}%
\makeatother
\begin{document}

\preprint{AIP/123-QED}

\title{Anisotropic Electron Heating in an Electron Cyclotron Resonance Thruster with Magnetic Nozzle}
% Force line breaks with \\

\author{J. Porto}
\email{jcportoh@gmail.com}
\affiliation{Physics - Instrumentation and Space Department, ONERA/DPHY, Université Paris Saclay\\ F-91123 Palaiseau -- France.}
\affiliation{Sorbonne Université, Observatoire de Paris, PSL Research University, LERMA, CNRS UMR 8112\\ 75005 Paris -- France.}

\author{P.Q. Elias}%
\email{paul-quentin.elias@onera.fr}
\affiliation{Physics - Instrumentation and Space Department, ONERA/DPHY, Université Paris Saclay\\ F-91123 Palaiseau -- France.}

\author{A. Ciardi}%
\affiliation{Sorbonne Université, Observatoire de Paris, PSL Research University, LERMA, CNRS UMR 8112\\ 75005 Paris -- France.}

\date{\today}% It is always \today, today,

\begin{abstract}
In a grid-less Electron Cyclotron Resonance (ECR) plasma thruster with a diverging magnetic nozzle, the magnitude of the ambipolar field accelerating the positive ions depends of the perpendicular energy gained by the electrons.
This work investigates the heating of the electrons by electromagnetic waves, taking their bouncing motion into account in a confining well formed by the magnetic mirror force and the electrostatic potential of the thruster. An electromagnetic Particle-In-Cell (PIC) code is used to simulate the plasma in a magnetic field tube. The code's Maxwell solver is based on a semi-Lagrangian scheme known as the Constrained Interpolation Profile (CIP) which enables larger time steps. The results show that anisotropic plasma heating takes place exclusively inside the coaxial chamber, along a Doppler-broadened zone. It is also shown that a trapped population of electrons with a larger perpendicular energy exists in the plume.

\end{abstract}

\maketitle

\section{\label{sec:level1}Introduction}

Electric thrusters play a fundamental role in the field of space propulsion. Their main advantage lies in an efficient use of the propellant mass, and therefore a reduced consumption of propellant. Hall Effect Thrusters or Gridded Ion Engines are examples of the most well-known and flight-proven technologies in the current propulsion market nowadays. Both technologies eject an ion beam which is subsequently neutralized to prevent the spacecraft from charging. Several components of these technologies, such as the acceleration grid or the neutralizer, are subject to erosion and wear and for this reason, meeting the challenging lifetime targets requires careful optimization and demanding testing \cite{holste_ion_2020}. The complexity of some of the components has driven the investigation of alternative concepts of propulsion devices that require neither a grid nor a neutralizer. The Electron Cyclotron Resonance (ECR) plasma thruster \cite{Jarrige2013, Vialis2017a} is one of these concepts and it is the subject of the present study.

\begin{figure}[t]
     \centering
     \begin{subfigure}[b]{0.5\textwidth}
         \centering
         \includegraphics[width=8.0cm]{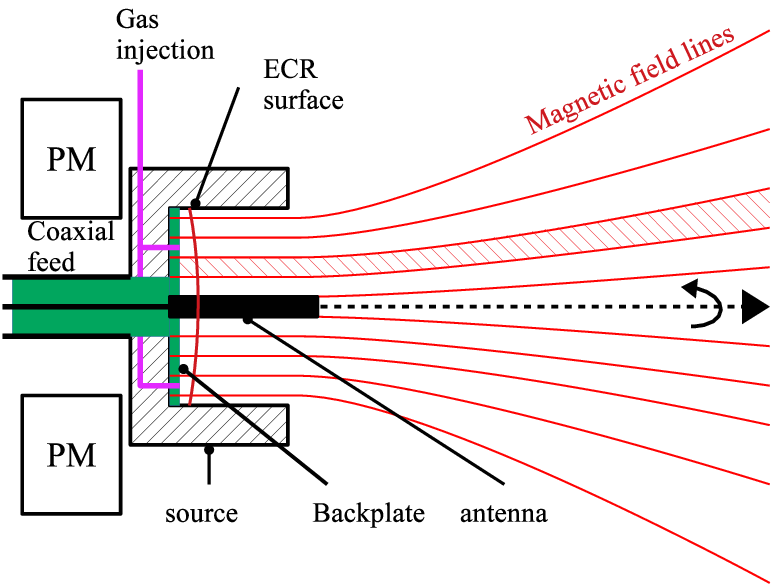}
         \vspace{0.1cm}
         \caption{}
         \label{ch4:fig:real_ecr}
     \end{subfigure}
     \vskip\baselineskip
     \vspace{-0.0cm}
     \captionsetup[subfigure]{margin={+0.0cm,0cm}}
     \begin{subfigure}[b]{0.5\textwidth}
         \centering
         \includegraphics[width=8.0cm]{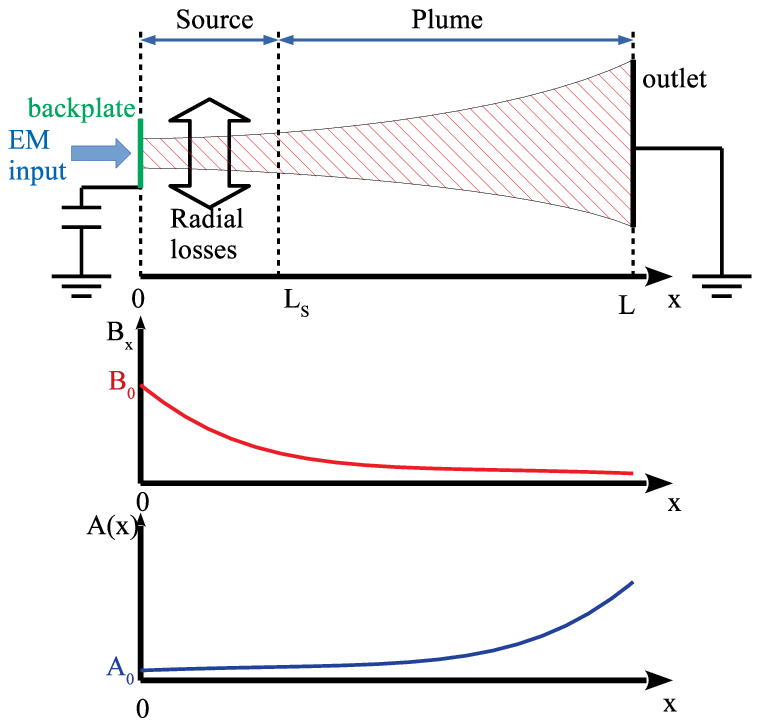}
         \caption{}
         \label{ch4:fig:modele}
     \end{subfigure}
        % \centering
         \caption{ECR thruster: (a) Schematic view of the coaxial source. The magnetic field lines are shown in red. The dashed surface corresponds to the flux tube (b) Schematic view flux tube used for the quasi-1D model. The exit plane of the coaxial source of length $L_S$ is represented by the dashed line. The end of the computational domain is reached at $x=L$. The axial magnetic profile and tube cross section along the axis are shown in red and blue, respectively.}
        \label{ch4:fig:real_and_model}
\end{figure}

The ECR plasma thruster consists of a semi-open chamber where a quasi-neutral plasma is heated by electron cyclotron resonant microwaves at $\SI{2.45}{GHz}$, and accelerated by a magnetic nozzle. This concept was first proposed in the 1960s in the works of \citet{Miller1963} and \citet{Nagatomo}, then further developed by \citet{sercel_experimental_1993}. These studies used a prototype with a wave-guide structure to couple the microwaves to the plasma. Their results showed that it was possible to achieve specific impulses and thrust values high enough to be of interest for space propulsion applications \cite{sercel_experimental_1993}. Nonetheless, the inefficiency, size and weight of the micro-wave sources and electromagnets at that time led to a stagnation of the research on ECR thrusters for several years. Interest for this technology arose again recently with experimental works \citep{Jarrige2013a, Jarrige2013c}. In particular, the use of coaxial microwave coupling structures and compact rare-earth permanent magnets were instrumental in designing compact sources (a schematic of the design is shown in Fig. \ref{ch4:fig:real_ecr}).

More experimental and theoretical efforts has since been made in order to get a deeper understanding of the physical phenomena governing the plasma heating and acceleration in the thruster. Experimental characterizations of the plasma properties have been carried out using different measurement techniques such as Langmuir and Faraday probes, Laser Induced Fluorescence diagnostics, diamagnetic loops and thrust balances \cite{Jarrige2013,hepner_low_2018, sara2019,correyero_characterization_2019,Vialis2018}. Unfortunately, most of the experimental studies so far have been limited to survey the plasma outside the thruster coaxial chamber. Recently, a resonant probe was developed to measure an electron density of about $\SI{1e11} cm ^{-3}$ at the source exit plane, close to the coaxial chamber \citep{Boni2021}. In the source, it is likely that the plasma density is higher ($\sim \SI{1e12} cm ^{-3}$)  with electron temperatures of a few tens of $eV$. 

From a theoretical point of view, as a first step, global models describing the energy balance in the plasma source were proposed as a means to obtain the key parameters of the thruster \cite{Cannat2015,verma_thrust_2020}. While this approach yielded good agreement with measured electron temperature at high mass flow rate or high pressure, they failed at the lower mass-flow rate where the thruster achieves its best performance. Indeed, the assumptions of uniform electron temperature and isotropic Maxwellian electron distribution are too crude approximations when collisionality decreases and the electron mean free path becomes much larger than the source length: in that range non-local effects become prevalent, as electrons undergo a bouncing motion along the magnetic field line. Those electrons which cross the ECR surface can gain energy depending on their phase in the gyromotion\cite{lieberman_theory_1973}, which leads to a strong anisotropy of the distribution function.  An attempt to account for this stochastic heating in the plasma was made by considering the electron heating as a random walk in phase space \cite{Peterschmitt2019}. While this model provided a qualitative agreement with the measured ion energies, it could not account for the plasma feedback on the waves (assumed constant) and the collisions along the bouncing motion. Recently \citet{Alvaro2021} performed 2D axisymmetric simulations of the thruster with a hybrid model consisting of particle-in-cell (PIC) ions and a fluid model for the mass-less electrons. One of the main findings of this study was the identification of different regions in the source where the waves are either propagating or evanescent, with most of the power absorption taking place close to the inner conductor, near the ECR surface. By acting as a sink for the plasma, the inner conductor induces a decrease of the plasma density in its vicinity, enabling the propagation of electromagnetic waves downstream of the ECR surface.  These features lead to the formation of a hot electron beam close to the inner conductor, with a colder plasma in the bulk of the source. While these 2D results provided important insights on the operation of the thruster, some assumptions of the fluid model limit the validity of the results obtained from these simulations. The most important one being the assumption of isotropic electron temperature which excludes anisotropic heating in the directions parallel and perpendicular to the magnetostatic field.

This latter point is still an open question for this technology. Indeed, ECR heating is expected to lead to anisotropic heating of the electron translation modes. This difference affects the power losses near the source walls and the potential drop in the magnetic nozzle \cite{Hooper1995}. However, most of the electron temperature measurements performed in the thruster plume did not differentiate perpendicular and parallel electron temperature (with respect to the local magnetic field direction). A way to measure the electron temperature anisotropy is, for example, incoherent Thomson scattering\cite{vincent_compact_2018}, but this type of measurement is not presently available in the ECR source. At any rate, this heating is intimately linked to the absorption of the electromagnetic waves in the coaxial source.

Another issue is the non-local transport due to the bouncing magnetized electrons in the nozzle (the electron mean free path is greater than the source radius). In particular, the production and the heating of the electrons are not necessarily at the same location.

While gaining a better understanding of these issues should firstly rely on experimental measurements, the challenges associated with such an investigation are a strong incentive to use numerical models, even if simplified, to investigate the main physical processes at play in ECR thrusters. In particular, such a model should be able to account for the self-consistent wave absorption and the anistropic heating, as well as the non-local effects and bouncing motion of the particles. Electromagnetic kinetic models, such as Particle-In-Cell (PIC) or Vlasov methods, are natural candidates for this task.

There are currently a few works using kinetic simulations of propulsion devices exploiting the ECR phenomenon, however the majority of these developments are concerned with gridded ion thrusters with ECR heating \cite{Nishiyama2006,Shen2017,Takao2014,Yamashita2019}, where the plasma acceleration is achieved by a grid-imposed electric field and not the plasma expansion in the magnetic nozzle, as in our design. \citet{Takao2014} successfully modelled a gridded ion thruster where the ions are produced in an ECR source at 4.2 GHz. The authors used a Particle-In-Cell (PIC) code considering the microwave electric field as a temporal modulation of its initial amplitude obtained by simulating the microwave propagation without plasma. Therefore, in this approach the plasma feedback on the wave was considered negligible.

The main purpose of the our study is to perform full-PIC electromagnetic simulations of the plasma in the thruster, taking into account the plasma feedback on the wave propagation, and to investigate the heating and confinement of the electrons. For this purpose it is necessary to simulate the microwave propagation and its interaction with the charged particles in the source and the nozzle region. However, due to the complexity and computational cost of simulating a full 3D configuration (which should include the nozzle region), we restrict our investigation to the simplified case of an isolated magnetic flux tube. This approximation effectively restricts the phase space to 4 dimensions (1 dimension in space, 3 dimensions in velocity space), and a 1D3V electromagnetic Particle-In-Cell can be used to model the ECR thruster. 

We show that the electron heating takes place over a broad region in the thruster source and leads to a significant anisotropy (ratio $T_{e\perp}/T_{e\parallel}\in [2.5,7.5]$). The perpendicular electron temperature reaches a first maximum in the source and, surprisingly, has a second maximum in the downstream region. The explanation for these features lies in the confinement of electrons in the potential well formed by the combination of the diverging magnetic field and the electrostatic potential.

\section{\label{num_model} Numerical Model}

\subsection{\label{ch4:sec:numerical_model1d}The Quasi-One-Dimensional Approach}

In the coaxial ECR thruster, an axially magnetized cylindrical permanent magnet creates a diverging static magnetic field $\mathbf{B}_{MS}$ in the source and in the plume region \cite{Cannat2015, Vialis2018}. This shape for the magnetostatic field was chosen to ensure a magnetic confinement at the close end of the coaxial chamber (called backplate in Fig. \ref{ch4:fig:real_and_model}) while allowing the electrons to get accelerated in the plume thanks to the divergence of the magnetic field lines.

In fact, the ECR uses a diverging magnetic field whose magnitude decreases from approximately \SI{100}{mT} at the back of the source to around \SI{5}{mT} \SI{10}{\centi\meter} downstream of the thruster exit plane. Under these conditions, assuming an electron temperature around $T_e\simeq \SI{10}{eV}$, the Larmor radius of the electrons is between $r_L\simeq \SI{0.07}{mm}-\SI{1.4}{mm}$. Thus electrons are strongly magnetized in the source and in the near-field plume region, while ions remain mostly unmagnetized. As a consequence, before the onset of plasma detachment, electrons and ions are bound to the magnetic field tube. Several mechanisms may account for the plasma detachment\cite{ahedo_plasma_2011} : collisions, stretching of the magnetic field lines, electron demagnetization and plume instabilities. While it is out of the scope of this work to study the dominant mechanisms, several recent works have investigated some of these effect in 2D PIC simulations\cite{andrewsFullyKineticModel2022,huKineticInsightsThrust2021,emotoNumericalInvestigationInternal2021}.In this work, we decided to rely on experimental evidence to define the section of the nozzle where the plasma \textit{remains} bound to the field lines. Recently, \citet{little_electron_2019} have mapped the plasma potential in a magnetic nozzle to show that a good criterion for detachment is $\chi_p=r_L / L_{\nabla B}\simeq 0.1$, where $r_L$ is the electron Larmor radius and $L_{\nabla B}=(\nabla B / B)^{-1}$ is the characteristic length scale of the magnetic field gradient. In the region of the nozzle where $\chi_p=r_L / L_{\nabla B}< 0.1$, the plasma remains attached to the magnetic field. In our case, we considered a magnetic field with $L_{\nabla B}\simeq 5-\SI{10}{cm}$. Under this condition, we have $\chi_p < 0.1$ up to $L=\SI{10}{cm}$ downstream of the nozzle, and it is a reasonable assumption to consider that electrons do not detach from the magnetic field tube over this distance.

As a consequence of this assumption, we decided to consider the creation and formation of the plasma enclosed in a magnetic field tube of length $L=\SI{10}{cm}$. More precisely, a portion of the thruster chamber and plume was represented by a quasi-1D model of a magnetic field tube with a varying cross-sectional area, as seen in Fig. \ref{ch4:fig:modele}. There are several examples of the use of quasi-1D models in the space propulsion field. \citet{Niewood1992} used it to model a Magnetoplasmadynamic thruster, while \citet{DeGiorgi2019} studied a Vaporizing Liquid Microthruster with this approach. Recently, \citet{saini_double_2020} also used this approach to model plasma expansion in a Radio-Frequency thruster. The moderate computational cost of a 1D3V model of the thruster facilitates the analysis of the plasma behavior in both the coaxial chamber and in the magnetic nozzle, and importantly, taking into account the nozzle is critical to resolve the bouncing motion of the electrons.

The quasi-1D model assumes that the electrons and the ions are confined within a diverging magnetic flux tube, whose area is related to the axial magnetic field intensity through the conservation of the magnetic flux:
    	\begin{equation}
            \label{ch4:eq:b_a_relationship}
            A(x)B_x(x) = A_{0} B_{0}
        \end{equation}
The model further assumes that the electromagnetic fields and all the plasma properties are constant across the section of the flux tube. For the ECR thruster under consideration \cite{Cannat2015, Vialis2018}, the shape of the magnetic field lines close to the antenna is well approximated by an exponential function. For the sake of simplicity we approximated the magnetic field as:
\begin{equation}
\label{eq:b_profile}
    B_x(x) = B_0 \exp(-\frac{x}{L_B})
\end{equation}
In addition, we considered cylindrical symmetry for the static magnetic field around the field tube centerline.

These assumptions mean that the particles guiding centers remain on the centerline. Since the plasma is assumed uniform in the cross section, this approach does not allow the formation of a diamagnetic current and $E\times B$ drifts.

From now on, the term \textit{parallel} and the subscript $\parallel$ will refer to the direction parallel to the magnetostatic field lines. Similarly, \textit{perpendicular} and the subscript $\perp$ refer to the direction perpendicular to the magnetostatic field lines. The source region, which corresponds to the coaxial cavity in Fig. \ref{ch4:fig:modele}, was defined by $0\leq x\leq L_S$, where $L_S$ is the coaxial source length. The plume region, which corresponds to the plasma expansion in vacuum, was defined by $x\geq L_S$.

\subsection{\label{code} Particle-In-Cell Code overview}

The simulations were carried out with the Particle-In-Cell (PIC) code \textit{Rhei}, which was developed to simulate low pressure cold plasmas and is adapted to parallel architectures. It can be run with either a pure MPI or a hybrid MPI/OpenMp parallelization. The code integrates a Monte-Carlo Collision (MCC) module to simulate the collisions between the charged particles and a prescribed neutral background. At each iteration, once the electrostatic and the electromagnetic fields were computed, the position of each macro particle labeled “$p$” was updated using $d\mathbf{x_p}/dt  = \mathbf{v_p}$, and the velocity using Eq. \ref{ch2:eq:motion}. Each macro-particle represents $W$ physical particles. The value of $W$ used in the simulation is given in table \ref{ch4:tab:parameters}.

\noindent
\begin{equation}
\label{ch2:eq:motion}
m_s \dv{\mathbf{v_p}}{t} = q_s \left [ \mathbf{E}_{ES_p} +\mathbf{E}_{EM_p} + \mathbf{v_p}\times \left (  \mathbf{B}_{MS_p}+\mathbf{B}_{EM_p} \right )\right ]
\end{equation}

In Eq. \ref{ch2:eq:motion}, $q_s$ is the charge of the particle, $m_s$ the mass, $\mathbf{x_p}$ the position, and $\mathbf{v_p}$ the velocity. Regarding the fields, they were computed at the location of the particle $p$ using linear interpolation function, where $\mathbf{E}_{ES_p}$ is electrostatic field from the charge distribution, $\mathbf{B}_{MS_p}$ is magnetostatic field from the permanent magnets and $\mathbf{E}_{EM_p}$ and $\mathbf{B}_{EM_p}$ are electromagnetic fields produced by the microwave source and by the plasma itself.

The equations of motion were integrated using the leap-frog method and the Boris scheme to get the $\mathbf{v} \times \mathbf{B}$ rotation from the Lorentz force\cite{Birdsall1991}. Details of the integration in the context of the quasi-1D model are provided in appendix \ref{annexe:motion}. Particle quantities were projected on a uniform grid using linear shape functions.

The Rhei code development follows a \textit{test-driven} approach to ensure the robustness and the maintainability of the code over time. Additionally, several test cases were run as a validation of the code. The first elementary test was the simulation of a magnetic bottle. The simulation domain, with a converging-diverging parabolic magnetic field, was uniformly loaded with a Maxwellian electron population. At the end of the simulation the electron distribution in velocity space $v_{\parallel},v_{\perp}$ was plotted to verify that the loss cone angle is coherent with the expected theoretical value $\arcsin(\sqrt{B_0/B_{Max}})$. The second elementary test concerned the electromagnetic modes in a one-dimensional magnetized plasma. The simulation domain was initialized with a uniform Maxwellian distribution of electrons and cold ions. The random fluctuations excited the modes of the plasma. The resulting dispersion curves were obtained by computing the discrete 2D Fourier transform of the electric fields during the simulation. This was compared to the expected theoretical description of the extraordinary and the ordinary wave. Finally, the third test case was the classical capacitively coupled discharge in Helium, which verified in particular the collision module\cite{Turner2013}.

\subsubsection{Collisions}\label{sec:collision}

The Monte-Carlo Collision module used the \textit{Null Collision} technique \citep{Skullerud1968} to speed-up the computation of the collisions by removing the velocity dependency of the total collision cross-section. Assuming $N_p$ collision processes defined by their respective cross sections $\sigma_i(v), i=1..N_p$, a null collision cross-section is defined as $\sigma_0(v)$ such that:
\begin{equation}
    \sigma_0(v)=\max_{v\geq 0} \left (\sum_{i=1}^{N_p} \sigma_i(v) \right) -\sum_{i=1}^{N_p} \sigma_i(v) 
\end{equation}

A first test over all the particles of species $s$ found the fraction of particles which undergo a collision with the background. In that case the total cross section $\sigma^T=\sum_{i=0}{N_p}\sigma_i(v)$ (including the null collision process)  did not depend on the velocity (thus avoiding a costly interpolation to get the cross section for all the particles). Then a second test among those selected particles computed all the collision cross sections for their given relative velocity and determined which cross section to use (including the null collision). When this test pointed to the null-collision cross section, then the particle did not experience an actual collision and was left unchanged. When the test pointed to another cross section, the the particles experienced a collision.

For the collisions of electrons with Xenon neutrals, we considered a simplified set of three processes: elastic, ionization, and excitation. Excitation processes were lumped into a single process. Electron impact ionization and excitation were taken from the Morgan (Kinema Research \& Software) database, while the total elastic scattering is from Ref. \onlinecite{mceachran_momentum_2014}. For all electronic processes, we assumed an isotropic scattering of the relative velocity vector between the electron and the target during the collision. For the ionization collisions, the kinetic energy of the projectile electron was equally split (after subtracting the threshold energy) between the secondaries. For the collisions of Xenon ions with Xenon neutrals, we considered isotropic scattering and backscattering\cite{Turner2013}. Ion cross sections comes from Ref. \onlinecite{piscitelli_ion_2003}. All electronic and ionic processes conserved momentum and total energy (kinetic plus internal). In order to start with a simplified description simulating weakly ionized plasmas, in which the collisions with the neutral particles are the dominant process, Coulomb collisions were not considered in the code. Indeed, the electron-ion collision frequency $\nu_{ei}$, for Maxwellian electrons, is given by:
\begin{equation}
    \nu_{ei}=\frac{\omega_p}{\Lambda_{ei}}\ln{\Lambda_{ei}}
\end{equation}
Here, $\omega_p$ is the plasma frequency and $\ln\Lambda$ is the Coulomb logarithm. For the typical simulation conditions in the thruster source, as it will be shown below, the maximum plasma density was $n_e\sim \SI{1e11}{\per\cubic\centi\meter}$, the electron temperature was $T_e\sim\SI{10}{eV}$ and the electron-neutral elastic collision frequency was $\nu_{en}\sim \SI{1e7}{\per\second}$. This gave $\ln{\Lambda}\sim 12-15$, $\omega_p\sim \SI{1.8e10}{\radian\per\second}$. Consequently, the maximum electron-ion collision frequency was $\nu_{ei}\sim \SI{1e5}{\per\second}$, much less than the the electron-neutral collision frequency $\nu_{en}$. 

The neutral gas in the thruster is injected at the backplate (see Fig. \ref{ch4:fig:real_ecr}) and expands in the source resulting in a decreasing density. Since there is no measurement of the neutral gas density profile in the thruster, and to avoid a costly particle simulation of the neutral particles, we modelled this expansion heuristically by assuming that the neutral background density followed an exponential profile:
\noindent
\begin{equation}
    \label{ch2:eq:neutrals}
    n_{n}(x)=n_{n_{0}}\exp\left(-\frac{x}{L_n}\right )
\end{equation}

where $n_{n_{0}}$ is the maximum density of neutrals found at the close end of the source, and $L_{n}$ is the neutrals density characteristic length. The assumption of a time-independent neutral gas density profile means that the simulation did not conserve the total mass, momentum and energy. In addition, it means that the neutral gas depletion due to ionizing collisions was not considered. However, both of those limitations are acceptable in the frame of this work which does not seek to compute the total thrust and energy balance but is concerned with the particle heating and trapping. To estimate the neutral depletion we note that the ion removal is driven by their velocity (at most $\SI{10}{\kilo\meter\per\second}$), while the neutral removal is driven by their thermal speed, $\sim \SI{200}{\meter\per\second}$. From mass balance, the neutral inflow is balanced by the ion flux and the neutral outflow. Using the characteristic speeds and the typical parameters for the gas density $n_g\sim \SI{1e14}{\per\cubic\centi\meter}$, and plasma density $n_e^{max}\sim\SI{3e11}{\per\cubic\centi\meter}$, gives a neutral depletion of at most 10\%, indicating that the assumption of a static background remained consistent with the assumed density profile.

\subsubsection{Fields solvers}
The \textit{Rhei} code solved the Poisson equation to compute the electrostatic space potential $\Phi$ and the electric field ($\mathbf{E}_{ES}=E_x\mathbf{x}$) using Eq. \ref{ch2:eq:poisson} where the charge density is $\rho_{s}$. The solver implements a second order finite difference discretization and the resulting linear system is inverted using an iterative method (GMRES)\cite{stoer_iterative_2002}.

\noindent
\begin{equation}
    \nabla^2 \Phi  (\mathbf{x},t) =  -\frac{{\rho_s}(\mathbf{x},t)}{\epsilon}
    \label{ch2:eq:poisson}
\end{equation}

In addition, an electromagnetic solver computed the fields produced by the microwave source and by the plasma itself: $\mathbf{E}_{EM}=E_y\mathbf{y}+E_z\mathbf{z}$ and $\mathbf{B}_{EM}=B_y\mathbf{y}+B_z\mathbf{z}$. This solver was based on the Constrained Interpolation Profile (CIP) method explained in detail in Ref. \onlinecite{Porto2021}. This method considers not only the electromagnetic fields but also their spatial derivatives, therefore suppressing instabilities and providing lower numerical dispersion even when using coarse grids and large time steps \citep{Kajita2014}. The use of this method is a novel solution for a PIC code since most of the electromagnetic solvers are based on conventional approaches like the finite-difference time-domain method (FDTD). It was shown that it provides higher accuracy than the latter under the condition of identical cell size\citep{Okubo2007_bis}.

The CIP method is a semi-Lagrangian scheme that circumvents the Courant-Friedrichs-Lewy (CFL) stability condition\citep{Yabe2001,P.K.Smolarkiewicz1992}, i.e., $(u \Delta t/ \Delta x) < 1$ where $u$ is the magnitude of the velocity, $\Delta t$ is the time-step, and $\Delta x$ the length interval. This feature allows computations with CFL values $\geq 1.0$, as can be seen in Ref. \onlinecite{Nie2018} and \onlinecite{Tachioka2012} where the authors performed simulations using a CFL value of $2.6$ in a Cartesian coordinate system. The gain in computational time, that is afforded by using high CFL values, is a key factor that enables the self-consistent kinetic simulations presented here to reach steady-state. In this paper, CFL values close to 3 were used for the simulations. As a check, simulations were also run with CFL=0.6 and compared to the results obtained with larger time steps. The results were identical to the one at larger time-steps, within small variations due to the noise inherent to the statistical nature of the PIC simulations.

Finally, the CIP scheme does not necessarily maintain the divergence-free condition for the dynamic field $\mathbf{B}_{EM}$. However, $\mathbf{B}_{EM}$ is smaller than the magnetostatic field (which has divergence equal to zero by construction, see appendix \ref{annexe:motion}) by several orders of magnitude, over the whole computational domain. Therefore, the resulting error on the total divergence was considered to be negligible.

\subsubsection{Boundary conditions}

As it was shown in Fig. \ref{ch4:fig:modele} when describing the model, the domain goes from $x=0$ at the left side which corresponds to the backplate and the microwave input, to the right-end at $x = L$, as discussed in \ref{ch4:sec:numerical_model1d}.

\textbf{Electrostatic}: At the right end of the computational domain $x = L$, we imposed a Dirichlet boundary condition, with $\Phi(L)=0$, to simulate the presence of a grounded vacuum chamber wall. The dielectric backplate, at $x = 0$, is in contact with the plasma and therefore its surface voltage $\Phi_{BP}$ is changed by the collection of charged particles. This can be modeled as a capacitor. Hence, the evolution of $\Phi_{BP}$ is given by $\Delta \Phi_{BP} =\Delta Q/(C \Delta t)$, where $\Delta Q$ is the charge deposited at the backplate at each time step, and $C$ is an equivalent capacitance under the assumption that the backplate is in contact with a grounded conductor. This capacitance is computed by considering that the backplate is a plane capacitor, its value is of a few picoFarads. Changing its magnitude modifies the charging rate of the backplate and thus the transient phase of the computation. However, its does not affect the steady-state voltage of the backplate. This approach guarantees that at steady-state, the ion flux equates the electron flux on the backplate. In principle the steady-state value of the backplate potential is also affected by other processes such as secondary electron emission or charge migration. However, for this study, these processes were neglected.

\textbf{Electromagnetic}: In the coaxial ECR thruster, the microwaves are injected as Transverse Electro-Magnetic (TEM) mode. For this 1D simulation, the TEM mode can be seen as a linearly polarized wave, where the radial component of the electric field is along the transverse $y$ axis, the azimuthal magnetic field defines the $z$ axis and the wavevector direction is along the longitudinal $x$ axis. Therefore, the microwaves were injected at the backplate as a propagating wave with a linear polarization along the \textit{y-axis}. The incident wave was parametrized by its power per unit area $P_{in}$ and its frequency $f_{EM}=\omega / 2\pi$. The electric fields from the injected linearly polarized wave were computed as $E_{y} = \sqrt{\mu c P_{in}} \sin{(\omega t)}$ and $E_z = 0$.

The injected microwave input power per unit area $P_{in}$ could be fixed, or it could be adapted to keep a roughly constant pre-defined number of particles $N_{target}$ during the transient phase. This feature was intended to speed up the simulations by reproducing a faster plasma response to a given variation in the simulation's parameters. The value of $P_{in}$ can be regulated with an attenuation factor $\alpha\leq 1$ varying with the number of particles in the domain: $\alpha = exp(- N_{particles} / N_{target})$. A run performed without this regulation confirmed that it did not have an effect on the final steady state but only on the duration of the transient phase.

\textbf{Particles}: We imposed a \textit{loss} condition at both ends of the domain, for both ions and electrons. Particles crossing these boundaries are suppressed from the simulation. As a simplifying assumption, secondary emission processes on the backplate were not considered in this first approach. 

\subsubsection{Cross field diffusion loss model}
\label{sec:crossfield}
Electron cross-field diffusion is an important mechanism to model to get a more accurate representation of the discharge loss mechanisms. Previous works using PIC codes for electric thrusters took it into account as wall losses by artificially increasing the collision rate or by using a profile of the cross field diffusion based on empirical evidence \citep{Blateau2000,Fox2003}. The electron balance equation is:
\noindent
\begin{equation}
    \dfrac{\partial n_e(\mathbf{r},t)}{\partial t}+\nabla_{\perp}\cdot n_e \mathbf{u_\perp}+\nabla_{\parallel}\cdot n_e \mathbf{u_\parallel}=k_{ion} n_e(\mathbf{r},t)
\end{equation}
Where $u_\perp$ and $u_\parallel$ are the electron macroscopic velocity perpendicular and parallel to the local magnetic field, respectively, and $k_{ion}$ is the ionization rate.

For our 1D3V simulations, the transport along the magnetic field is taken into account by the kinetic model. However, the perpendicular transport cannot be modeled with a 1D model. Therefore  we simulated the particle losses into the coaxial chamber walls using a phenomenological, Monte Carlo loss model, as shown in Fig. \ref{ch4:fig:modele}. The probability of an electron impacting the walls of the coaxial chamber was calculated from the diffusion equation of electrons across the magnetic field based on the assumption that their number density profile in the radial direction was independent of time and axial position. In a cylindrical coordinate system it can be expressed as the product  $n_{e}(x,r,t) = n_{e_{0}}(x,t)g(r)$. The balance equation (for a constant diffusion coefficient $D$) integrated over the radius of the flux tube $r_{max}$ was then given by:
\noindent
\begin{equation}
    \dfrac{\partial n_{e_0}(x,t)}{\partial t} +\dfrac{\partial n_{e_0}(x,t)u_x(x,t)}{\partial x}= - \nu_L n_{e_0}(x,t) + k_{ion} n_{e_0}(x,t) 
    \label{ch4:eq:loss_module}
\end{equation}
\noindent

With the loss frequency given by: 
\noindent
\begin{equation}
\label{ch4:eq:loss_freq}
\nu_{L} = -r_{max}\frac{g^\prime(r_{max})}{S} D
\end{equation}

Where $r_{max}$ is the radius of the flux tube, and the weighted cross section $S$ is given by:
\noindent
\begin{equation}
\label{ch4:eq:s_eq}
S = \int_0^{r_{max}} r g(r)dr \quad \text{and} 
\end{equation}

A first choice for the diffusion coefficient $D$ would be a coefficient based on classical diffusion obtained from theories on standard electron-neutral collisions. It can be seen in Eq. \ref{ch4:eq:d_classical} where $\tau = 1/ \nu$ is the collision period with the neutral background. However, the electron mobility tends to be higher than the value predicted by this classical diffusion approach \citep{Chen2016}. The cause of this discrepancy is an active area of research in the electric propulsion field \citep{Garrigues2009, Croes2017}. As a consequence, we decided to use the Bohm coefficient, which is a phenomenological coefficient accounting for the anomalous cross-field diffusion.

\noindent
\begin{equation}
\label{ch4:eq:d_classical}
D_{Bohm} = \frac{1}{16}\frac{k_{B}T_{e}}{eB} \quad \text{or} \quad D_{classical} = \frac{\omega_c \tau}{1+ (\omega_c \tau)^2}\frac{k_{B}T_{e}}{eB}
\end{equation}

The probability for a given particle to be lost between $t$ and $t+\Delta t$ is given by $p_{L}=\nu_{L} \Delta t$. In our quasi-1D model, the flux conservation relates the magnetostatic field to the cross-sectional area of the magnetic field tube as shown in Eq. \ref{ch4:eq:b_a_relationship}. In this work we assumed $g(r)=J_0(k_0 r / r_{max})$, with $k_0$ the first zero of the Bessel function. Then Eq. \ref{ch4:eq:loss_freq} has the form $\nu_L \propto k_0^2/r_{max}^2 D$. Since $D\propto B^{-1}$ and in the model $B S=B\pi r_{max}^2$ is a constant, the loss probability does not depend on the position along the flux tube and is given by:

\noindent
\begin{equation}
\label{ch4:eq:prob_loss}
p_{L}(x) = \frac{2}{3}\frac{\pi}{16} k_0^2 \left(\frac{1}{2}m \langle v(x)^2 \rangle \right) \frac{dt}{eA_{0}B_{0}}
\end{equation}

Where $\Delta t$ is the time step, $m\langle v(x)^2 \rangle/2$ is the electron's mean kinetic energy, and $A_{0}=A(0)$ is the cross-section of the magnetic field tube at $x=0$. The losses are computed at each time-step. For all electrons in the source (such as $x\leq L_S$, $L_S$ being the length of the coaxial chamber as shown in Fig. \ref{ch4:fig:modele}), the probability $p_L$ is computed using equation \ref{ch4:eq:prob_loss}. A random number $x$ is drawn from a uniform distribution. If $x\leq p_L$, the electron and a neighboring ion are removed from the simulation.

\subsection{\label{set_up}Simulation Setup}

The electron dynamics and the electromagnetic solver were updated every iteration. For these one-dimensional calculations the real Xenon mass for the ions was used and to speed up the calculations a subcycling was used so the ion's position and velocity that were updated every 10 time steps as given by $\Delta t_{ions}$ in Table \ref{ch4:tab:parameters}. The collisions were also computed every 10 time steps as given by $\Delta t_{coll}$. The charged particle's population was seeded using a uniform density distribution ($N\sim 10^3$). The electron's initial energy along each of the \textit{x-y-z} axis was set to $T_e =\SI{20}{eV}$, while ions were assumed cold $T_i=\SI{0.03}{eV}$. These values were intended to reproduce a non-equilibrium plasma at low density. The choice of the initial electron temperature  $T_e=\SI{20}{eV}$ is somewhat arbitrary. Checks run with several energy values between \SI{10}{eV} and \SI{30}{eV} showed no impact of the initial electron energy on the final characteristic of the steady state. To sustain the plasma at the beginning, a plasma source located at \SI{2}{mm} from the backplate injected electrons at \SI{3e5}{\meter\per\second} and ions at \SI{3e2}{\meter\per\second} during the first \SI{150}{\nano\second} of the simulation. These velocities were specified along each of the \textit{x-y-z} axis. Here the idea was to sustain the initial plasma long enough for the ionization to pick up and the plasma density to grow. The conditions for the simulation presented below are shown in table \ref{ch4:tab:parameters}. With this choice of magnetic field profile, the resonance condition $f_{EM}=eB/ 2\pi m$ was met at $x=\SI{6.7}{mm}$.

\begin{table}[h]
\centering
\caption{Simulation parameters for the electromagnetic full PIC simulations using the quasi-one-dimensional model.}
\label{ch4:tab:parameters}
\begin{tabular}[t]{|c|c|c|}
\hline
\textbf{Parameter} & \textbf{Description} & \textbf{Value}\\
\hline
\hline
$\Delta t$ & Time step & 1.6 ps \\
\hline
$\Delta x$ & Mesh spacing & \SI{167}{\micro\meter} \\
\hline
$C$ & CFL condition & 2.87 \\
\hline
$f_{_{EM}}$ & Microwave frequency & 2.45 GHz \\
\hline
$L_{S}$ & Coaxial chamber length & 20 mm \\
\hline
$x_{_{ECR}}$ & ECR surface position & 6.7 mm \\
\hline
W & Weight for the charged particles &  $2\times 10^5$ \\
\hline
$L_{D}$ & Computational domain length & 100 mm \\
\hline
$n_{n_{0}}$ & Maximum number density of neutrals & 
$\SI{8e19}{} \text{m}^{-3}$ \\
\hline
$L_{n}$ & Neutral density characteristic length & 1.0 cm \\
\hline
$A_{L}$ & Cross-sectional area for the loss module & \begin{math}\SI{1}{\centi\meter} ^{2} \end{math} \\
\hline
$\Delta t_{ions}$ & Time step to push the ions & 10$\Delta t$ \\
\hline
$\Delta t_{coll}$ & Time step for collisions & 10$\Delta t$ \\
\hline
\end{tabular}
\end{table}%

\begin{figure}[h]
     \centering
     \includegraphics[width=8cm]{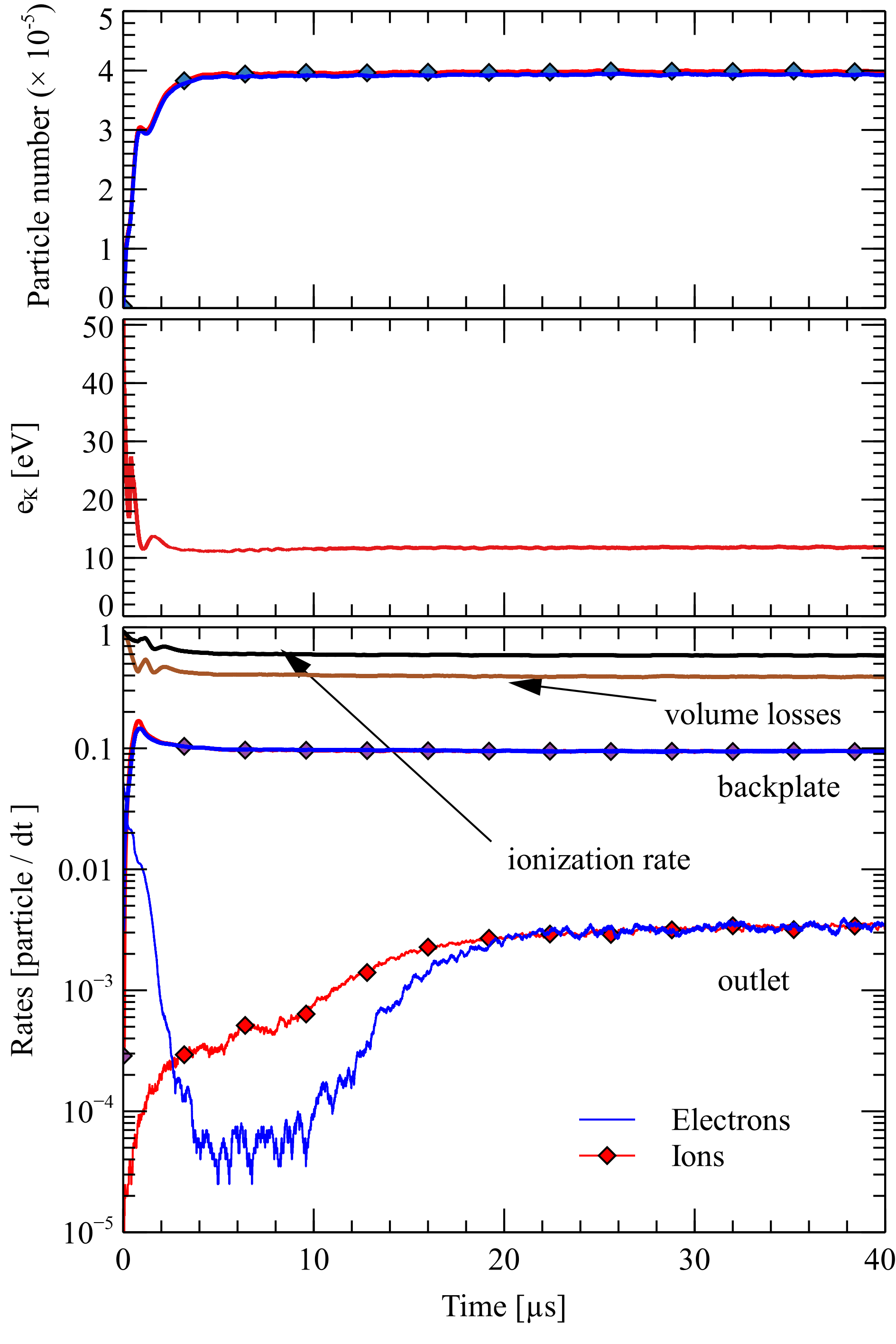}
     \caption{Time evolution of simulation quantities. Top frame : total number of macro-particles (ions and electrons) in the simulation. Middle frame : Mean kinetic energy of the electrons. Bottom frame : particle fluxes at the boundaries (backplate and outlet) and particle source and sink terms in the whole computational domain. The volume loss gives the average number of particle lost per timestep due to the Bohm loss model. For the steady-state analysis, the particles quantities are sampled after $t=\SI{30}{\micro \second}$}
     \label{fig:time_evolution}
 \end{figure}

The simulation was run until it reached a steady state, usually after around $\SI{30}{\micro \second}$ which represents between 5 to 8 ion transit times. The definition of this steady state was done by following up the variation of the total number of particles in the domain, its mean kinetic energy, and the particle flux at the backplate and the plume since an equal number of ions and electrons must be impacting both surfaces, as shown in Fig. \ref{fig:time_evolution}. At the end of the simulation, when the steady state was reached, the plasma properties were obtained by calculating the time average for each parameter over several time steps. Overall, the wall time of the simulation was 44 hours, with 12 OpenMP threads.

\section{\label{results}Results}

\begin{figure}[htbp]
     \centering
     \includegraphics[width=8cm]{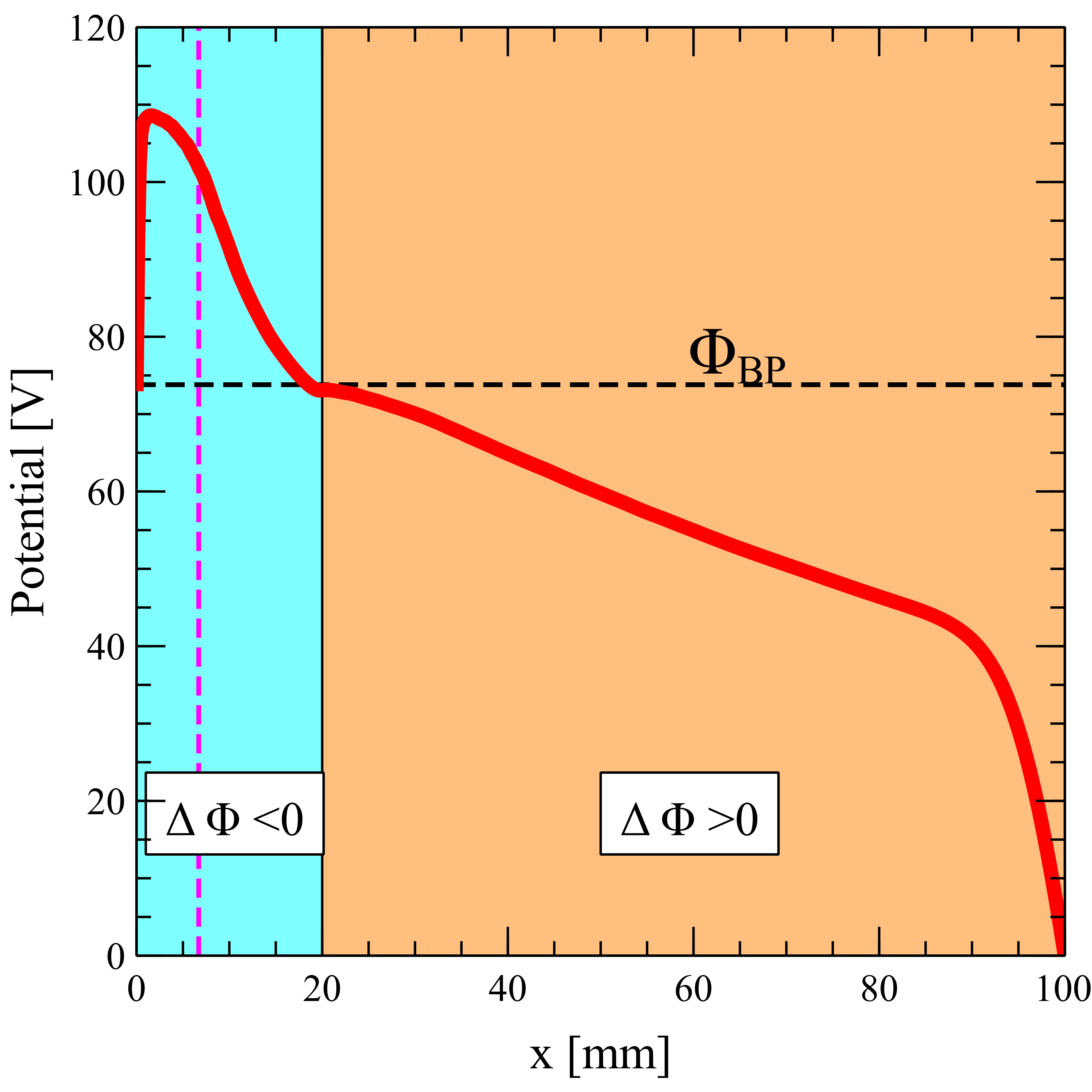}
     \caption{Plasma potential. The vertical dashed line indicates the ECR surface location. The horizontal dashed line shows the backplate potential $\Phi_{BP}$. The two colored zones delineate the regions where the plasma potential is above ($\Delta\Phi>>0$ or below ($\Delta\Phi<0$) the backplate potential.}
     \label{ch4:fig:profile_potentiel_plasma}
 \end{figure}

\begin{figure}[htbp]
     \centering
     \includegraphics[width=8cm]{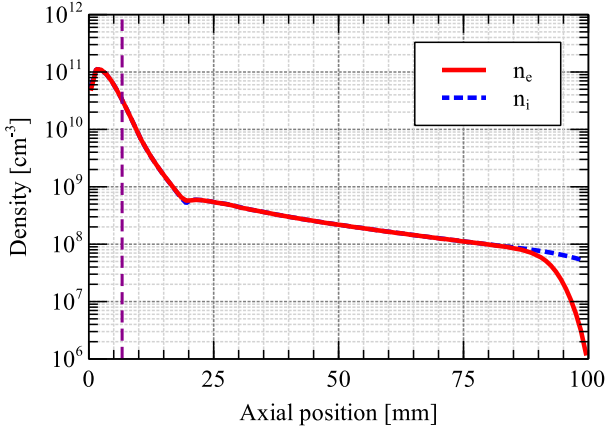}
     \caption{Electron (solid) and ion (dashed) number densities. The dashed line indicates the ECR surface location.}
     \label{ch4:fig:profile_densite}
 \end{figure}

Figure \ref{ch4:fig:profile_potentiel_plasma} shows the steady-state plasma potential distribution over the whole computational domain. Except for slight random fluctuations on the instantaneous potential, no large scale fluctuations were observed. Time-averaging improved the signal to noise ratio but did not blur the shape of the profile. The backplate reached a positive steady state potential of around $\SI{70}{\volt}$. The peak of the plasma potential was $\SI{105}{\volt}$, and it was reached at around \SI{3}{mm}, interestingly not at the ECR surface (indicated with a vertical dashed line). Indeed, the shapes of the plasma density and potential are driven by the ionization rate. For this simulation, the ionization rate was monotonically decreasing, because the background density decrease was faster than the ionization rate increase due to the plasma heating. As a consequence, the maximum ionization was upstream of the ECR surface. This  peak in the plasma potential formed a barrier. As a result, ions collected on the backplate were necessarily created in a region where $x\leq\SI{3}{mm}$, while ions collected downstream were created in a region where $x\geq\SI{3}{mm}$ and accelerated into the nozzle by the ambipolar electric field. Sheaths were formed at the backplate and the vacuum chamber wall. The plasma sheath width on the backplate was $\sim \SI{0.1}{mm}$. At the downstream end $x=L$ of the domain, as shown by Fig. \ref{ch4:fig:profile_densite}, the plasma sheath began at around \SI{90}{mm}. This size was consistent with a Debye length $\lambda_D\sim 1-\SI{5}{mm}$ for a plasma density around $\SI{1e8}{\per \cubic \centi \meter}$. The electron and ion peak number density was $\SI{1.12e11}{\per\cubic \centi\meter}$ at $x=\SI{1.5}{mm}$. Recall that the ECR condition is met at $x=\SI{6.7}{mm}$.

\begin{figure}[t]
\centering
\includegraphics[width=8cm]{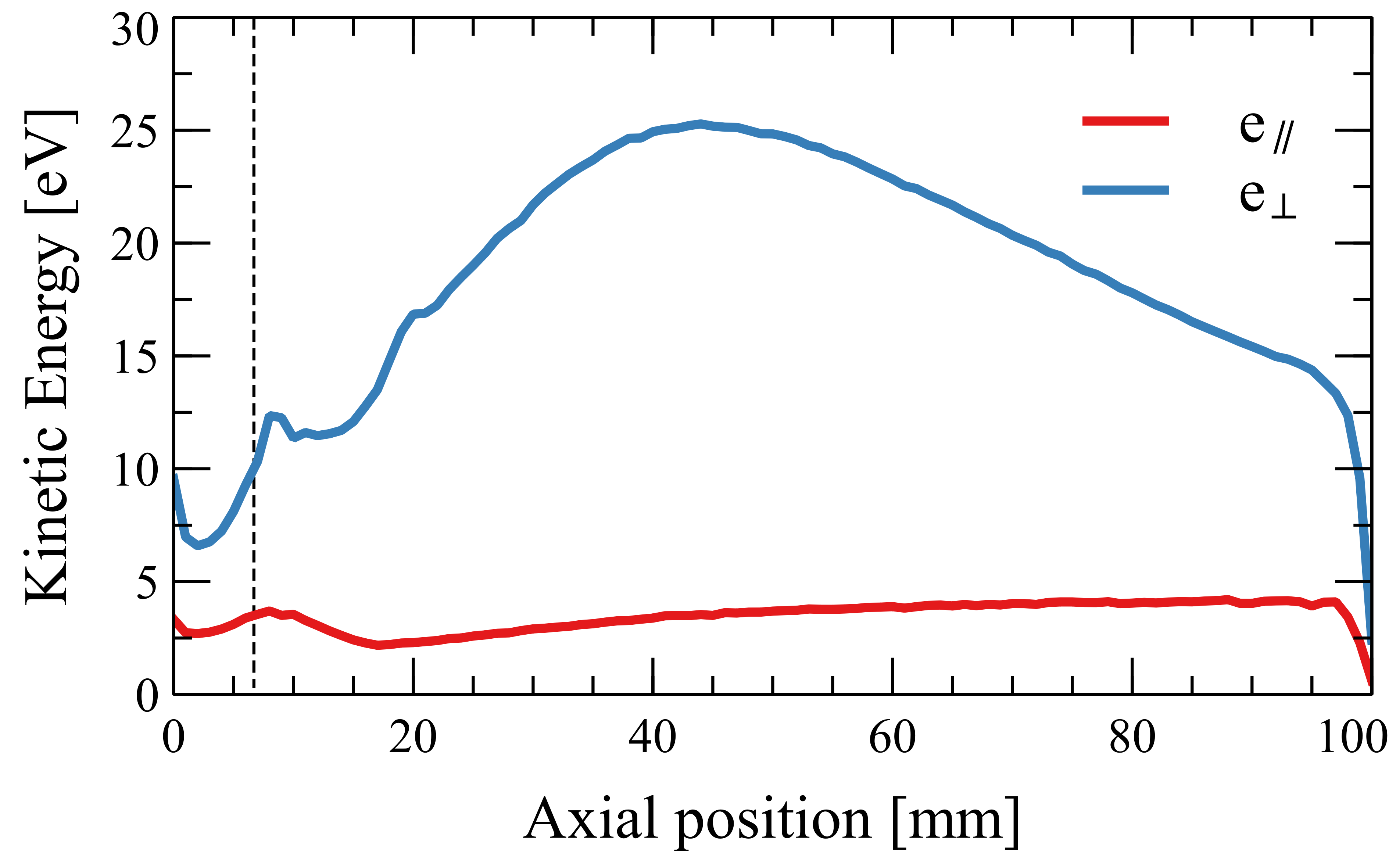}
\caption{Electron's mean kinetic energies: $e_\parallel$ longitudinal (blue line) and $e_\perp$ perpendicular (red line) directions. The location of the ECR surface is shown by the dashed line.}
\label{ch4:fig:mean_Ek}
\end{figure}

In Fig. \ref{ch4:fig:mean_Ek} we plotted the electron's mean kinetic energy in both the axial ($e_{\parallel}$) and the perpendicular ($e_{\perp}$) direction as a function of the axial position on the domain. First, we observed that the mean parallel kinetic energy remained nearly constant, around $4-\SI{5}{eV}$, over the whole simulation domain. The perpendicular energy was higher than the parallel component, which underlined the anisotropic heating of the electrons in this thruster. More precisely, the mean perpendicular kinetic energy $e_{\perp}$ reached a first peak at around $x=\SI{9}{mm}$ and then decreased before reaching a global maximum of \SI{25}{eV} at $x = \SI{45}{mm}$. After this point, $e_{\perp}$ decreased until the end of the simulation domain. Over the whole simulation domain, the anisotropy ratio $T_{e,\perp}/T_{e,\parallel}$ was found to vary between 2.5 and 7.5. Given that the ECR heating increases the perpendicular energy of the electrons, it was expected to see an anisotropic behavior depending on the direction parallel or perpendicular to the magnetic field lines. However, the second broad energy peak in the downstream part of the magnetic nozzle was puzzling. To get a better understanding of these feature, it was necessary to evaluate the energy deposition by the electromagnetic field.

\medskip

\subsection{ Electromagnetic Energy deposition in the source}
\label{ch4:subsec:electron_heating}

To understand how the field energy was transferred to the particles, we considered the energy balance equation, including the Poynting flux (its derivation is provided in appendix \ref{annexe:energyequation}). 
\begin{equation}
    \pdv{\epsilon_{EM}+\epsilon}{t}+\nabla \cdot (\mathbf{Q}+\Pi) = S_{coll}
\end{equation}
In this equation, $\epsilon$ and $\epsilon_{EM}$ stands for the electron kinetic energy density and the electromagnetic energy density, respectively. $\mathbf{Q}$ and $\Pi$ are the kinetic energy flux and electromagnetic energy flux; $S_{coll}$, whose expression is given in Eq. \ref{eq:scolle}, is the volume power loss term due to the collisions and the diffusion. This latter term account for the energy lost by elastic and inelastic collisions with the neutral background and by the particles removed by the loss model. Since we were interested in the steady state regime, and considering that the field quantities depend on $x$ only, this was further simplified to:
\begin{equation}
    \frac{1}{A}\pdv{}{x}A(Q_e+Q_i+\Pi)=S_{e,coll}+S_{i,coll}
    \label{eq:eqEN}
\end{equation}
where we separated the time-averaged total energy flux into a kinetic contribution due to the electrons $Q_e$ , the ions $Q_i$ and the electromagnetic contribution $\Pi$. The kinetic energy flux of the electron was further separated into a flux of parallel energy $Q_{e,\parallel}$ and perpendicular energy $Q_{e,\perp}$ (see appendix \ref{annexe:energyequation}).

To quantify the magnitude and direction of the energy exchanges between the electromagnetic field and the particles, the different terms of Eq. \ref{eq:eqEN} were evaluated. To do so, the particles and field quantities were sampled in the steady state phase (after $t=\SI{30}{\micro\second}$, see fig. \ref{fig:time_evolution}). First, particles were sorted in 120 spatial bins equally spaced along the axial direction. In each bin, the moments of particle distribution provided the total energy flux $\mathbf{Q}_e$ and $\mathbf{Q}_i$, as detailed in appendix \ref{annexe:energyequation}. Second, the cross product of the electric and magnetic field provided the axial component of the Poynting vector. This vector was time-averaged over a period corresponding to an integer number of wave periods.

\begin{figure}[htbp]
\centering
\includegraphics[width=8cm]{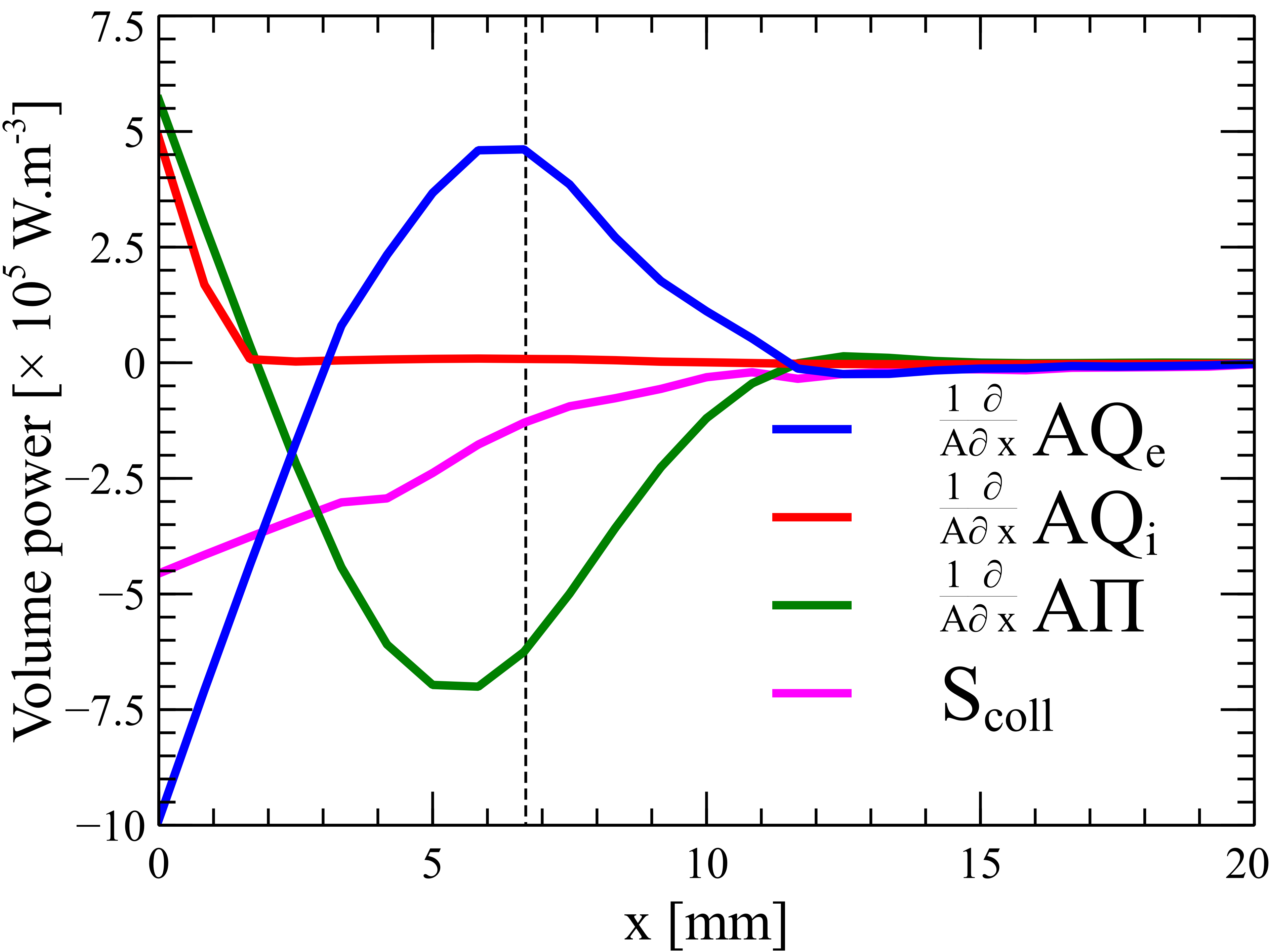}
\caption{Energy source terms of eq. \ref{eq:eqEN} along the axial direction. The location of the ECR surface is shown by the dashed line.}
\label{fig:source}
\end{figure}

\begin{figure}[htbp]
\centering
\includegraphics[width=8cm]{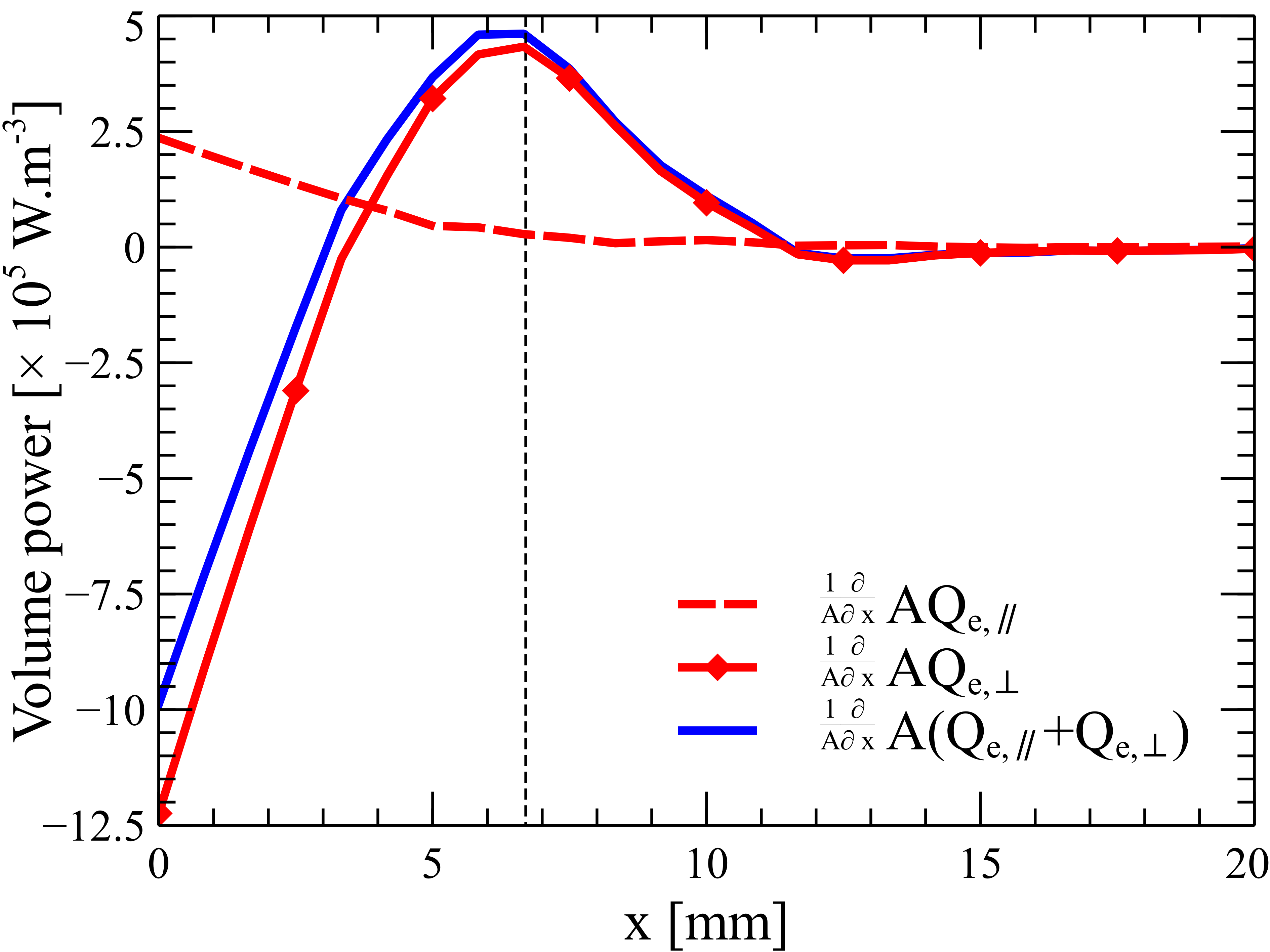}
\caption{Perpendicular and parallel energy source terms for the electrons along the axial direction. The location of the ECR surface is shown by the vertical dashed line.}
\label{fig:fluxQ}
\end{figure}

The results, plotted in Fig. \ref{fig:source}, showed first that the energy source terms are negligible in the plume region ($x\geq \SI{20}{\milli\meter}$) compared to the source region ($x < \SI{20}{\milli\meter}$). In the plume region, the magnitude of the source terms is below $\SI{1e4}{\watt \per \cubic \meter}$. For that magnitude, the signal is dominated by the statistical noise of the PIC simulation. This noise drowns the finer features of the source terms, especially for the electrons which are more prone to statistical noise. Nevertheless, this underlines that most of the energy exchanges take place in the source region and cannot explain the secondary peak for mean perpendicular kinetic energy $e_{\perp}$ observed in the plume region (Fig. \ref{ch4:fig:mean_Ek}). 

Second, the collision source term is negative over the whole source region. Its magnitude is maximum near the backplate, where the neutral and plasma densities are higher, and decreases along the source axis. This behaviour is not unexpected, since this term lumps together the contribution of inelastic collisions and the diffusion model: these two processes are loss mechanisms for the plasma. Given that the plasma density and the neutral gas density decrease as we move away from the backplate, the collision frequency drops and the magnitude of the energy loss decreases.

Third, the sign and magnitude of the source terms reveal different phenomena. For $x\leq \SI{3}{\milli\meter}$, the terms linked to the Poynting vector are positive, while the source term due to the electron energy flux is negative. This indicates an energy conversion from the electron kinetic energy to the field energy. In parallel, the ion source term is positive, which points to a gain of energy in the sheath. For the $\SI{3}{\milli\meter} < x \leq \SI{10}{\milli\meter}$ range, the Poynting source term shows a negative peak, while the source term due to the electron energy flux is positive, with a peak centered on the ECR surface location. In that case, there is a transfer of energy from the field to the electrons. This latter feature corresponds to the ECR heating of the perpendicular energy mode of the electrons. In fact, this appears when considering the parallel and perpendicular contributions to the source term in  Fig. \ref{fig:fluxQ}. The perpendicular source term dominates over the parallel source term, with a peak centered on the ECR surface. The axial extent of this peak shows that the perpendicular mode of the electrons is heated in a zone of about $\Delta x_{_{ECR}} \approx \SI{6}{mm}$, i.e., from $x = \SI{3}{mm}$ to $x = \SI{9}{mm}$. Considering that the ECR condition is only met on a specific surface, the presence of an extended region may seem surprising. However, as will be discussed later, most of the electrons in the magnetic field tube are confined and undergo a bouncing motion in the potential well formed by the electrostatic field and the magnetic mirror force. These bouncing electrons can cross the resonance surface with a significant parallel velocity, thus one may expect a shift of the resonance condition due to the Doppler effect. The width $\Delta x_{ECR}$ can be compared with the expected value for a Doppler broadened resonance $\Delta x_{D}$ in Eq. \ref{ch4:eq:doppler} \citep{Williamson1992}.

\noindent
\begin{equation}
    \label{ch4:eq:doppler}
    \Delta x_{_{D}} = \sqrt{\frac{2\pi v_{\parallel}}{\frac{\omega_c}{B_{_{MS}}} \abs{\pdv{B_{_{MS}}}{x}}}}
\end{equation}

Using the electrons' mean axial velocity $v_{\parallel}$ is $\SI{1.1e6}{\meter\per\second}$ we obtain $\Delta x_{_{D}} = \SI{4.7}{mm}$. However, as it will be shown later in Fig. \ref{ch4:fig:vpara_vperp}, there is a high dispersion for the values of $v_{\parallel}$. Therefore, we can expect a much larger Doppler broadening for the fastest electrons. The electrons' axial velocity can reach values up to $\SI{3.0e6}{\meter\per\second}$ around the ECR zone. With this velocity, we can compute a maximum Doppler broadening of $\SI{7.8}{mm}$, which means that $\SI{4.8}{ mm}< \Delta x_{_{D}} < \SI{7.8}{mm}$. The ECR heating zone obtained in the simulations is consistent with the one expected analytically, indicating that the Doppler effect is a good candidate to explain the width of the heating zone observed in Fig. \ref{fig:source}.

Because of Doppler-broadening, ECR heating of the electrons can even occur when the ECR surface is outside the plasma source. Indeed, the fact that the plasma in the thruster can be sustained even with an ECR zone outside the coaxial chamber was demonstrated experimentally by \citet{Vialis2018PhD}, where the location of the resonance surface was placed at $x=\SI{-0.17}{ mm}$ and $x=\SI{-0.77}{mm}$. Doppler broadening is a possible explanation to this finding and this hypothesis was tested in simulations using the same parameters described in Table \ref{ch4:tab:parameters} but with an input microwave frequency of $f_{_{EM}} =\SI{2.9}{GHz}$. It was possible to sustain a discharge for this frequency even though the ECR condition was meet at $x=\SI{-1.78}{ mm}$ (i.e., upstream of the plasma source).

In summary, the analysis of the power deposition shows that the energy transfer occurs mainly in the source region. The electrons absorb the wave energy over a Doppler-broadened volume. This energy goes preferentially to the perpendicular energy mode and leads to an anisotropy ratio $T_{e,\perp}/T_{e,\parallel}\in [2.5,7.5]$. This energy deposition from the field to the perpendicular energy mode can explain the first peak in perpendicular energy seen in Fig. \ref{ch4:fig:mean_Ek}. However, the source terms are negligible in the plume region and thus cannot explain the broad peak observed in this region.

\subsection{\label{ch4:subsec:electron_dynamics} Electron confinement in the magnetic nozzle}

\begin{figure}[!tbp]
    \captionsetup[subfigure]{margin={0.6cm,0cm}}
    \centering
    \begin{subfigure}[b]{0.45\textwidth}
        \centering
        \includegraphics[width=\textwidth]{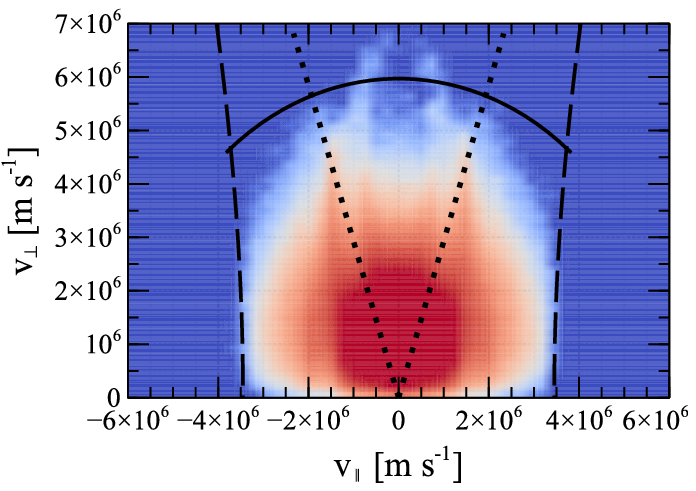}
        \caption[]%
        {{\small }} 
        \label{ch4:fig:vpara_vperp_a}
    \end{subfigure}
    \hfill
    \begin{subfigure}[b]{0.45\textwidth}  
        \centering 
        \includegraphics[width=\textwidth]{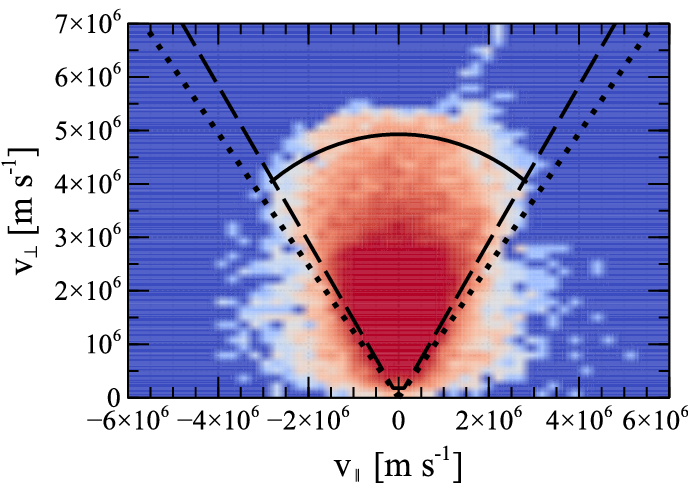}
        \caption[]%
        {{\small }}   
        \label{ch4:fig:vpara_vperp_c}
    \end{subfigure}
    \hfill
    \begin{subfigure}[b]{0.45\textwidth}  
        \centering 
        \includegraphics[width=\textwidth]{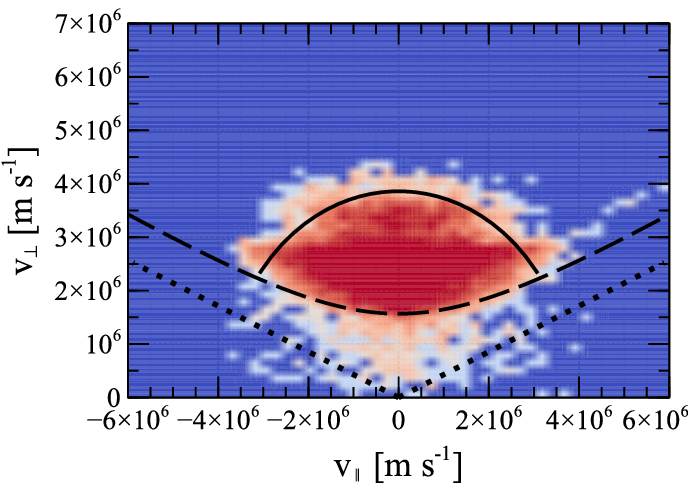}
        \caption[]%
        {{\small }} 
        \label{ch4:fig:vpara_vperp_f}
    \end{subfigure}
    \centering
    \caption{Electrons distribution in the velocity space $v_{\parallel},v_{\perp}$ plane at different locations on the simulation domain. The number of electrons has been normalized by the total number of electrons for each case independently. For each case, we also plotted what we call the \textit{confinement boundaries} described by an analytical model (Eq. \ref{ch4:eq:theta} and \ref{ch4:eq:confinement}). The dotted line is the magnetic confinement, the dashed line the electrostatic potential at the backplate, and the solid line is the electrostatic confinement on the plume. (a) x = 5 mm, (b) 20 mm, (c) x = 80 mm.}
    \label{ch4:fig:vpara_vperp}
\end{figure}

To determine the factors driving the evolution of the mean electron energy, we considered the electron distribution in the nozzle region. In Fig. \ref{ch4:fig:vpara_vperp} we plotted the normalized electron distribution in the velocity space $v_{\parallel},v_{\perp}$ plane. The distribution was plotted at different locations in the computational domain. The results in Fig. \ref{ch4:fig:vpara_vperp} show that as we move downstream into the nozzle, we see an increased electron population with high energies in the perpendicular direction. To understand this phenomenon, we must first get a more detailed description of the electron confinement, i.e., how they get trapped and under which conditions they can leave the thruster.

Three loss pathways are identified for the electrons produced in the source.
\begin{enumerate}
    \item Cross field losses : electrons can diffuse across the magnetic field, due to anomalous transport, collision, etc. This is modeled by the phenomenological cross-field diffusion model presented in section \ref{sec:crossfield}.
    \item Losses at the downstream of the nozzle  electrons which have a kinetic energy sufficient to overcome the confining potential well are lost, alongside ions accelerated by this same potential drop.
    \item Electrons than can overcome both the repelling potential of the plasma sheath at the close-end of the source and the mirror-force are collected on the dielectric plate and will contribute to its surface charge.
\end{enumerate}

While the first loss mechanism does not depend on the kinetic energy of a single electron but rather on the mean kinetic energy at a given location, the two other mechanisms depend on the electron kinetic energy. An electron moving along the magnetic field will see both an electrostatic potential $\Phi$ (Fig. \ref{ch4:fig:profile_potentiel_plasma}) and the magnetostatic field $B$. Depending on its initial kinetic and potential energies, its trajectory might have turning points (where $v_\parallel=0$) within the domain, or out of the domain. In the first case, this electron is confined in the potential well. In the latter case, the electron is lost, either downstream (loss pathway 2) or at the backplate (loss pathway 3). Now the limiting case between confined / unconfined electrons is when the turning points are located at the boundaries. This will define the necessary conditions for the electrons confinement. Let us note $\Phi_{BP}$ the potential of the backplate. Neglecting the plasma-wave interaction, and noting $\mu$ the magnetic moment, the equation for the total energy of an electron is:

\noindent
\begin{equation}
\label{ch4:eq:motion}
E_{total} = \frac{1}{2}m v_{\parallel}^2 + \mu B - e\Phi
\end{equation}

\begin{figure}[h]
\centering
\includegraphics[width=7cm]{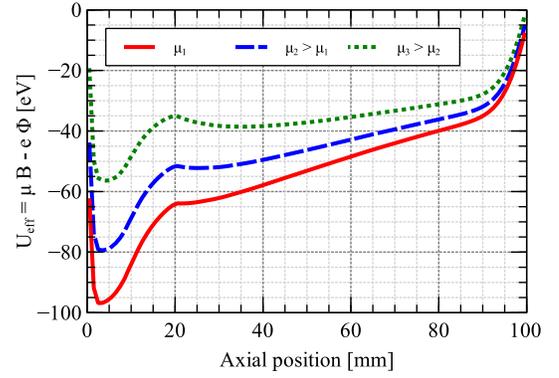}
\caption{Schematic view of the effective potential profile for arbitrary values of the magnetic moment.}
\label{ch1:fig:u_eff}
\end{figure}

From Eq. \ref{ch4:eq:motion} we can say that the electron is oscillating in an \textit{effective potential} given by $U_{eff} = \mu B -e \Phi$, where $\mu B$ represents the magnetic confinement as the electron moves towards the backplate while $-e \Phi$ is the electrostatic confinement given by the plasma potential. It can be seen in Fig. \ref{ch1:fig:u_eff} for different arbitrary values of the magnetic moment taken from the results of the simulation. Its concave shape explains the confinement of the electrons in the ECR thruster inside this potential well.

Let us now consider an electron moving from an arbitrary initial point to a turning point at a position $x_0$ along the longitudinal direction, such that $v_\parallel(x_0)=0$. The energy conservation between any initial location and the turning point gives:

\noindent
\begin{equation}
v_{\parallel}^2 + v_{\perp}^2 \left( 1 - \frac{B_{x_0}}{B}\right)  =  - \frac{2e}{m}\Delta \Phi
\label{ch4:eq:motion2}
\end{equation}

Where $\Delta \Phi = \Phi_{x_0}-\Phi$. Now let us consider the loss pathways 2 and 3 identified above.

For an electron lost to the downstream boundary (pathway 2), the confinement condition is obtained by setting $x_0=L$. In that case, given the divergence of the magnetic field, we have $0<1-B(L)/B<1$ and $\Delta \Phi<0$. Thus equation \ref{ch4:eq:motion2} describes an ellipse in the $v_\parallel-v_\perp$ plane. Electrons in the ellipse have there turning points $x_0\leq L$ and remain confined by the electrostatic well. Electrons out of the ellipse can overcome this electrostatic confinement and are lost downstream.

For an electron lost to the backplate (pathway 3), the confinement condition is obtained by setting $x_0=0$. Because the magnetic field is monotonically decreasing, $1-B(0)/B<0$. In addition, $\Phi(0)=\Phi_{BP}$. We define the loss cone angle as:

\noindent
\begin{equation}
\label{ch4:eq:theta}
sin(\theta) = \sqrt{\frac{B}{B(0)}}
\end{equation}
Equation \ref{ch4:eq:motion2} can be recast as:
\noindent
\begin{equation}
\label{ch4:eq:confinement}
v_{\perp}^2 = tan^2(\theta) \left( v_{\parallel}^2 + \frac{2e}{m}\Delta \Phi \right)
\end{equation}
Depending on the sign of $\Delta\Phi$, three cases are possible:

\begin{figure}%[h]
    %\captionsetup[subfigure]{margin={0.4cm,0cm}}
    %\centering
    \begin{subfigure}[b]{0.3\textwidth}   
        \centering 
        \includegraphics[width=\textwidth]{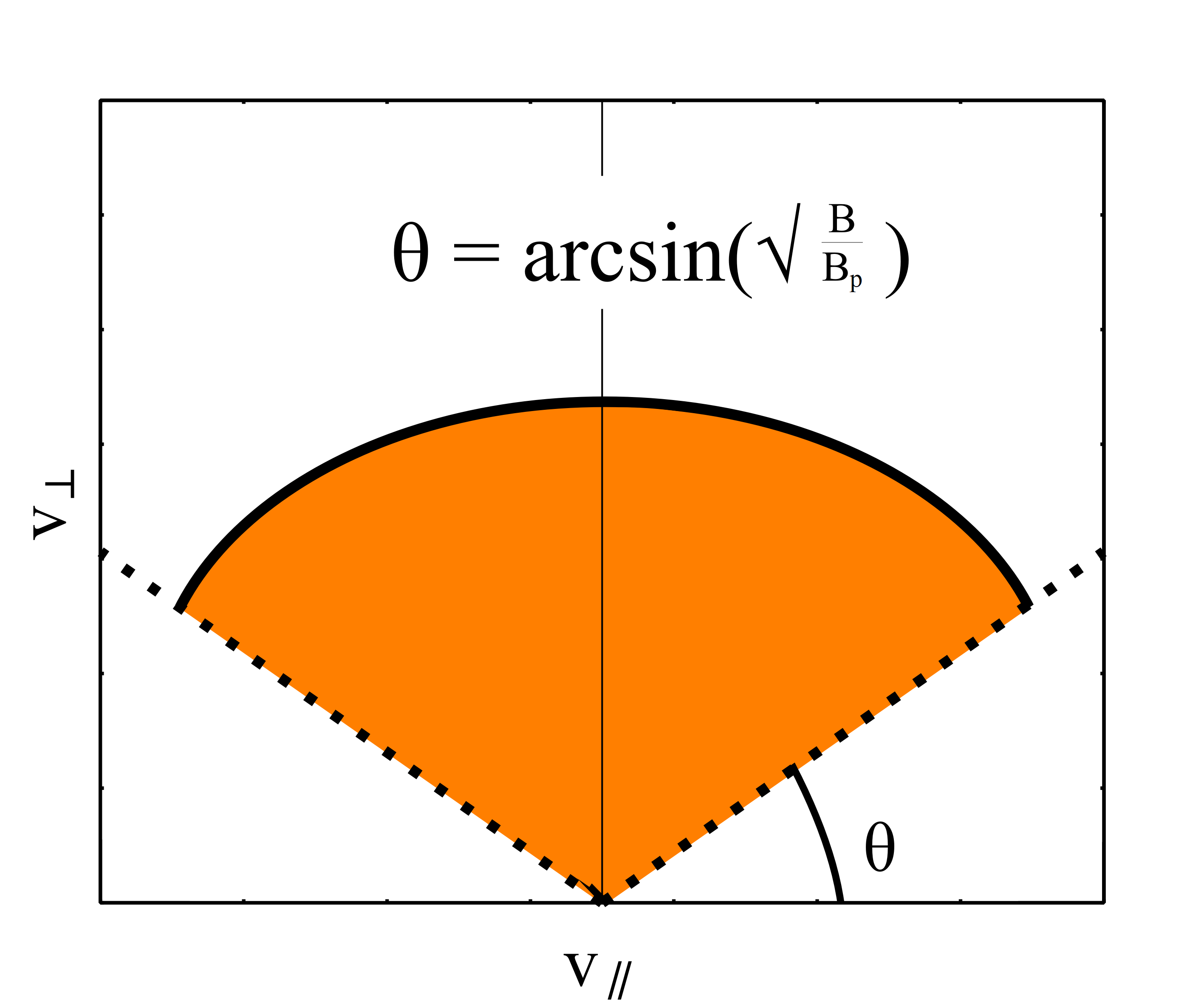}
        \caption[]%
        {{\small}}   
        \label{ch4:fig:confinement_theory_d}
    \end{subfigure}
    \vskip\baselineskip
    \vspace{-0.2cm}
    \begin{subfigure}[b]{0.3\textwidth}   
        \centering 
        \includegraphics[width=\textwidth]{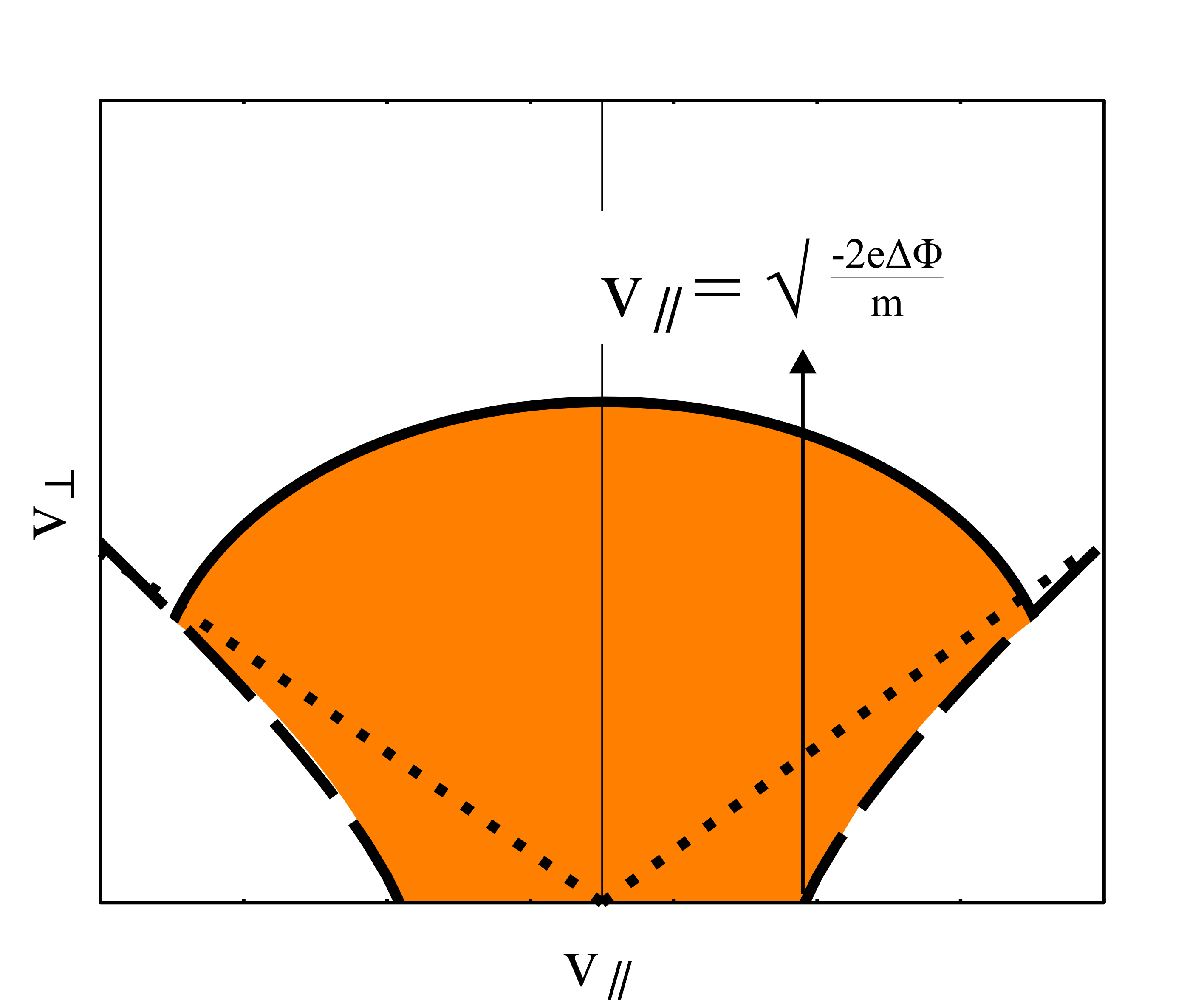}
        \caption[]%
        {{\small }}
        \label{ch4:fig:confinement_theory_e}
    \end{subfigure}
    \vskip\baselineskip
    \vspace{-0.2cm}
    \begin{subfigure}[b]{0.3\textwidth}  
        \centering 
        \includegraphics[width=\textwidth]{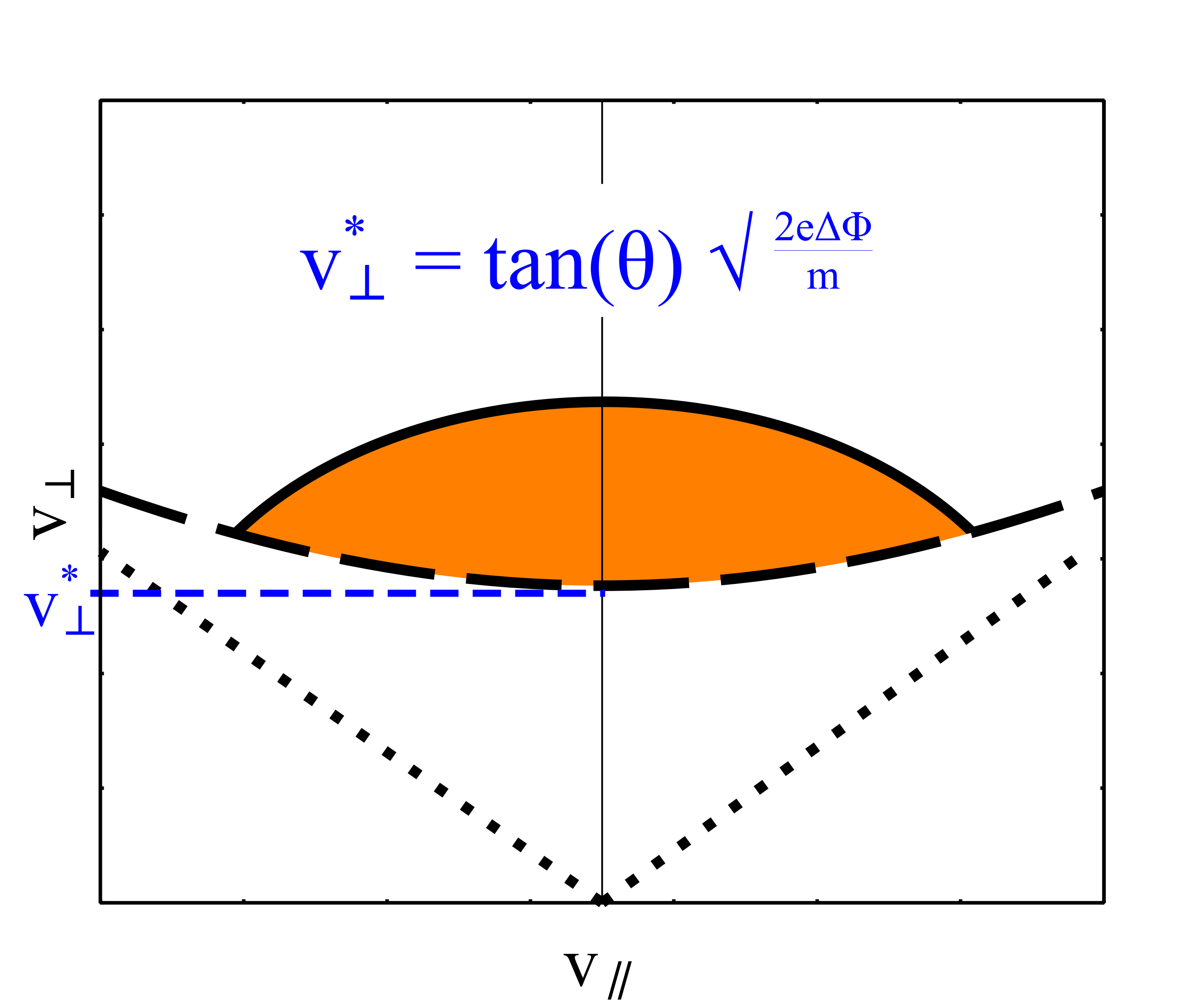}
        \caption[]%
        {{\small}}
        \label{ch4:fig:confinement_theory_f}
    \end{subfigure}
    \caption{Confinement boundaries in velocity space on the $v_{\parallel},v_{\perp}$ plane where the orange colored zones describe the electrons being trapped in the ECR thruster. Solid line for the plume electrostatic confinement, and dashed (electrostatic) plus dotted (magnetic) lines for the backplate confinement. Where: (a) $\Delta\Phi =0$, (b) $\Delta\Phi <0$, and (c) $\Delta\Phi >0$.}
    \label{ch4:fig:confinement_theory}
\end{figure}
% \twocolumngrid
% \end{widetext}

 \begin{itemize}
    \item $\Delta\Phi = \Phi_{BP}-\Phi = 0$: The confinement of the electron at the backplate is given exclusively for the topology of the magnetostatic field according to the loss cone angle $\theta$. Those electrons with values for $v_{\parallel},v_{\perp}$ such as they are located in the loss cone and will therefore be lost at the backplate (Fig. \ref{ch4:fig:confinement_theory_d}).

    \item $\Delta \Phi = \Phi_{BP}-\Phi < 0$: The backplate potential repels the negative charges and thus confines the electron. Consequently, a fraction of the electron in the loss cone will be reflected back and stay confined (Fig. \ref{ch4:fig:confinement_theory_e}).

    \item $\Delta \Phi = \Phi_{BP}-\Phi > 0$: The backplate potential attracts the electrons. Thus, even electrons out of the loss cone are collected. Therefore, the confined electrons are those that meet two conditions: they are not on the loss cone for the magnetic field, and they are energetic enough in the perpendicular direction to avoid being lost at the backplate thanks to the electrostatic acceleration towards it (Fig. \ref{ch4:fig:confinement_theory_f}).
    
 \end{itemize}
Fig. \ref{ch4:fig:profile_potentiel_plasma}, displays the sign of $\Delta\Phi$ in the nozzle. The combination of the loss conditions at the backplate (pathway 3) and downstream (pathway 2) delimits a confinement volume in phase space where electrons are confined, as shown in Fig. \ref{ch4:fig:confinement_theory}. Pitch-angle scattering, either caused by collisions with the neutral background or by the electromagnetic field, enables electrons to cross the confinement volume boundaries. Depending on which boundary is crossed, electrons are lost at the backplate or at the downstream side of the nozzle. Indeed, it is important to recall that the electron deconfinement is mainly driven by these two phenomena in the source region. Since the neutral background density decreases exponentially, most of the collisions occur in the source. In addition, as shown above, wave absorption happens over a few millimeters in the source. As a consequence, the electron deconfinement rate is driven by the wave interaction and the collisions:

\begin{itemize}
    \item  Interaction with the electromagnetic wave is akin to a scattering of the electron momentum \cite{kaganovichElectronBoltzmannKinetic2000}. After several passages through the ECR heating zone, the electron may gain enough energy to overcome the electrostatic barrier and escape into the plume.
    \item If the electron undergoes an elastic collision, it will randomly scatter its velocity vector. If the electron scattered momentum falls in the loss region defined by Eq. \ref{ch4:eq:confinement} and shown in Fig. \ref{ch4:fig:confinement_theory}, the particle is lost at the backplate.
\end{itemize}

If we now go back to the results in Fig. \ref{ch4:fig:vpara_vperp} for the electron distribution in the velocity space $v_{\parallel},v_{\perp}$ plane, we notice that as we move downstream into the plume, the confinement boundaries change. There is a transition from a confinement boundary as the one in Fig. \ref{ch4:fig:confinement_theory_e} ($\Delta \Phi < 0$) to the one in Fig. \ref{ch4:fig:confinement_theory_f} ($\Delta \Phi > 0$). This transition is a consequence of the fact that, as shown in Fig. \ref{ch4:fig:profile_potentiel_plasma} the plasma potential is greater than the backplate potential in the source $\Phi>\Phi_{BP}$ and lesser than $\Phi_{BP}$ downstream. Thus, inside the coaxial chamber, the plasma potential is such that $\Delta \Phi = \Phi_{BP}-\Phi < 0$ (Fig. \ref{ch4:fig:confinement_theory_e}), and in the plume section it is such that $\Delta \Phi = \Phi_{BP}-\Phi > 0$ (Fig. \ref{ch4:fig:confinement_theory_f}). As a consequence, as we move downstream into the magnetic nozzle, the mean perpendicular kinetic energy can increase. However, this is not given by an additional heating phase but as a result of confining only a highly energetic electron population in the perpendicular direction. Those electrons with a low perpendicular kinetic energy (i.e., below the dashed line) are lost at the backplate as previously described. It can be seen as a \textit{filtering process} where only the hot electrons are confined, and this is what we see when plotting the electron perpendicular kinetic energy in Fig. \ref{ch4:fig:mean_Ek}. Further downstream of the magnetic nozzle, the magnitude of the potential well to the end of the nozzle decreases, while the magnitude of the attracting potential drop to the backplate increases. This results in a narrower distribution for the confined population and finally a decrease in the mean perpendicular kinetic energy.

\section{Conclusions}

We have performed electromagnetic full-PIC simulations of the ECR thruster using a 1D3V model that allowed us to shed light onto some of its working principles. The results confirmed the expected anisotropic behavior for the electrons' energies in the direction perpendicular and parallel to the magnetic field lines and a peak for the mean perpendicular energy near the resonance zone. The microwave energy injected at the backplate of the thruster propagates through the coaxial chamber while being absorbed by the electrons increasing their kinetic energy perpendicular to the magnetic field lines. The absorption takes place exclusively inside the coaxial chamber on a zone of 6 mm around the resonance condition. This zone is coherent with the predicted value from  Doppler broadening. The width of this heating zone may explain why the thruster works even with a configuration in which the resonance condition is met outside the coaxial chamber\cite{Vialis2018PhD}. From a practical point of view, this feature improves the reliability of the thruster, since it means that the thruster can still operate even when the magnetostatic field decreases, for example due to excessive heating of the magnets.

The results also show, unexpectedly, that the electrons' mean perpendicular energy has a second peak in the plume due to the confinement of highly energetic electrons. The confinement is determined by the backplate's potential, the magnetostatic field, and the potential drop on the plume. As a consequence, there is a population of trapped electrons with significant perpendicular kinetic energy in the downstream region of the magnetic nozzle. The existence of doubly trapped electron population has been investigated using a kinetic model with a paraxial approximation similar to this work \cite{ahedoMacroscopicParametricStudy2020}. While this work was assuming the shape of the initial distribution function, it has also been seen in the case of an anistropic distribution that the perpendicular electron temperature could increase in the diverging part of the nozzle \cite{correyeroplazaPhysicsPlasmaPlumes2020}.
We can speculate that these high temperature electrons trapped in the plume could be important to drive some instabilities observed in the plume that are thought to enhance cross-field transport of electrons and thus play a role in the detachment. In particular, Lower Hybrid drift instabilities have been recently observed in diverging magnetic nozzle. In these experiments, diamagnetic drift $\mathbf{v}_D=\nabla p_{e\perp} \times\mathbf{B} / en B^2 $ was identified for the primary energy source for the instability \cite{hepner_wave-driven_2020}. The trapped electrons could enhance the radial gradient in perpendicular pressure and thus enhance the diamagnetic drift.\

\appendix

\section{Electron motion integration in the Quasi-1D model} \label{annexe:motion}

\begin{figure}[h]
\centering
\includegraphics[width=8cm]{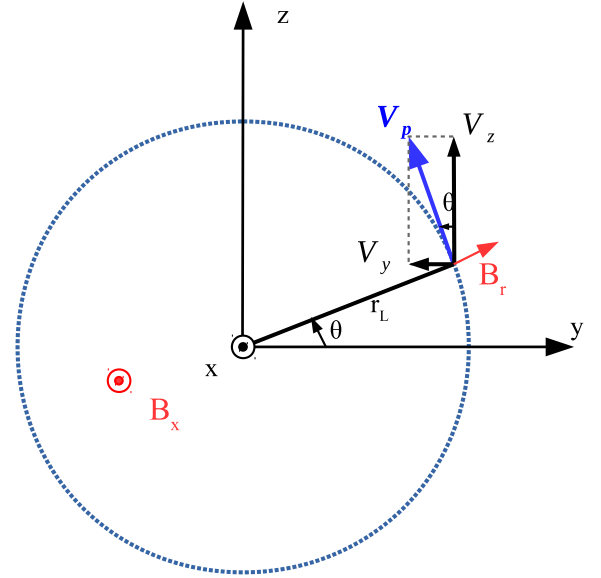}
\caption{Gyromotion in the plane normal to the axial static field $B_x$.}
\label{fig:circle}
\end{figure}

On one hand, the model assumes that the axial static magnetic field in the flux tube depends on $x$ only
\begin{equation}
    \frac{dB_x}{dx}=\alpha(x)
\end{equation}
In the quasi-1D model, the field remains constant across the section of the tube. 
On the other hand, in the ECR thruster the electromagnetic part of the magnetic field (the part due to the propagation of the electromagnetic power injected in the source) remains negligible compared to the static part. Indeed, assuming a locally plane wave, the Poynting vector $\mathbf{S}=\mathbf{E}\times\mathbf{B}/\mu_0$. Since $E\simeq c B$ and $S\simeq \SI{1}{\watt \per \square \centi \meter}$, the order of magnitude for the electromagnetic component of the magnetic field is $B\simeq \SI{1e-2}{\milli\tesla}$. The static part of the field is between \SI{10}{\milli\tesla} and \SI{10}{\milli\tesla}, much greater than the electromagnetic component. 
Therefore, using the divergence equation for the static magnetic field and neglecting the electromagnetic component, it is possible to obtain the radial component of the magnetic field :
\begin{equation}
    B_r(x,r) = -\frac{\alpha(x)r}{2}
    \label{eq:Br}
\end{equation}
Assuming this value for the radial part of the magnetic field ensures that the divergence condition is automatically enforced for the static part of the field. The electron guiding center is on the flux tube centerline. The Larmor radius of the electrons is given by:
\begin{equation}
    r_L(x)=\frac{V_{\perp}(x)}{\omega_c(x)}
    \label{eq:rl}
\end{equation}
Where $V_{\perp}=\sqrt{v_y^2+v_z^2}$, $\omega_c(x)=e B_x(x)/m_e$. The gyromotion of the electron is shown in Fig. \ref{fig:circle}. Thus, knowing the particle velocities it is possible to obtain the phase angle $\theta$ in its gyromotion, as given in Eq. \ref{eq:phase}.

\begin{eqnarray}
\cos(\theta) = & \frac{y}{r_L(x)}=\frac{v_z}{V_{\perp}} \\
\sin(\theta) = & \frac{z}{r_L(x)}=-\frac{v_y}{V_{\perp}} 
\label{eq:phase}
\end{eqnarray}
Knowing the phase angle, sine and cosine, the $B_y$ an $B_z$ components can be deduced from eqs. \ref{eq:Br} and \ref{eq:phase}.

\section{Energy equation} 
\label{annexe:energyequation}

For the energy equation for the particles we consider the second order moment of the Vlasov equation. Multiplying the Vlasov equation for the electrons by $\int m_{e} v^2/2$, we obtain:

\noindent
\begin{eqnarray}
\label{eq:b1}
\pdv{}{t}&& \int \frac{m_{e} v^2}{2} f_e d^3 \mathbf{v} + \pdv{}{\mathbf{x}} \cdot \int \frac{m_{e} v^2}{2} \mathbf{v} f_e d^3 \mathbf{v}  \\ && + q_e \int \frac{v^2}{2} (\mathbf{E + v \times B}) \cdot \pdv{f_e}{\mathbf{v}}  d^3 \mathbf{v} = \int \frac{m_{e} v^2}{2} \left ( \pdv{f_e}{t} \right)_{col} d^3\mathbf{v} \nonumber
\end{eqnarray}

Where the right hand side lumps the contribution of collisions and the loss model detailed in section \ref{sec:collision} and \ref{sec:crossfield}, respectively. Eq. \ref{eq:b1} can be rewritten as:

\noindent
\begin{eqnarray}
     \pdv{\epsilon_e}{t}+\nabla \cdot \mathbf{Q}_{e} =& -\mathbf{j_e} \cdot \mathbf{E} + S_{coll} \label{ch4:eq:fluxe}
     \mathbf{Q}_{e} =& \int \frac{m_{e} v^2}{2} \mathbf{v} f_e d^3\mathbf{v} \\
     \epsilon_e=&\int \frac{m_{e} v^2}{2} f_e d^3\mathbf{v} r \\
	S_{e,coll} =& \int \frac{m_{e} v^2}{2} \left ( \pdv{f_e}{t} \right)_{coll} d^3\mathbf{v} \label{eq:scolle}
\end{eqnarray}
The same procedure can be applied to the ions:
\noindent
\begin{eqnarray}
     \pdv{\epsilon_i}{t}+\nabla \cdot \mathbf{Q}_{i} =&- \mathbf{j_i} \cdot \mathbf{E} + S_{coll} \label{ch4:eq:fluxi}
	\mathbf{Q}_{i} =& \int \frac{M_{i} v^2}{2} \mathbf{v} f_i d^3\mathbf{v}  \\
	\epsilon_i=&\int \frac{M_{i} v^2}{2} f_i d^3\mathbf{v}  \\
	S_{i,coll} =& \int \frac{M_{i} v^2}{2} \left ( \pdv{f_i}{t} \right)_{coll} d^3\mathbf{v} 
\end{eqnarray}
Considering the electron population, the total heat flux can be written as:
\begin{equation}
    \mathbf{Q}_e=\mathbf{q_e}+\underline{\underline{\mathbf{P_e}}}\cdot \mathbf{u_e}+n_e\mathbf{u_e}(e_K+E_K)
\end{equation}
Where the density is given by $n_e=\int f_e d^3\mathbf{v}$ and the macroscopic velocity by $n_e\mathbf{u}_e=\int f_e \mathbf{v} d^3\mathbf{v}$. The random part of the velocity is $\mathbf{c}=\mathbf{v}-\mathbf{u_e}$  The different terms are then:
\begin{eqnarray}
    \mathbf{q_e} = \int \frac{m_e}{2}c^2\mathbf{c} d^3\mathbf{c} \\
    \underline{\underline{\mathbf{P_e}}} = \int m_e \mathbf{c}\mathbf{c} d^3\mathbf{c}\\
    E_K = \frac{1}{2}m_e u_e^2 \\
    e_K = \int \frac{m_e}{2}c^2 d^3\mathbf{c}
\end{eqnarray}
Considering the one-dimensional approximation, we are considering only the axial (parallel) part of the total heat flux. Thus, after averaging over a period of the incoming wave : $<...>=\frac{1}{T}\int ... dt$, we obtain:
\begin{eqnarray}
     \pdv{}{x}A(x) \left <q_{e\parallel} \right>+A\left <P_{e\parallel}u_\parallel +  P_{e\perp}u_{\perp}\right > \nonumber \\
     + A(x)\left <u_{e\parallel}+n_e u_{e;\parallel}(e_K+E_K)\right> \nonumber \\
     =- A(x)\left<\mathbf{j}_e\cdot\mathbf{E}\right> + A(x)\left< S_{e,coll}\right>
\end{eqnarray}
A similar expression can be written for the ions. if we make use of the Poynting theorem, we can relate the time averaged joule term to the divergence of the pointing flux $\Pi=\frac{\mathbf{E}\times \mathbf{B}}{\mu_0}$:
\begin{equation}
    A(x)\left < (\mathbf{j}_e+\mathbf{j}_i)\cdot \mathbf{E}\right>+ \pdv{}{x}\left < \Pi \right > =0
\end{equation}

If we drop the $<...>$ symbol for simplicity, recalling that all quantities are time-averaged, one obtains equation \eqref{eq:eqEN}.

% \nocite{*}
\bibliography{aipsamp}% Produces the bibliography via BibTeX.

\end{document}